\documentclass[12pt,preprint]{aastex}

\newcommand{\jcomp}{ $Journal$ $of$ $Computational$ $Physics$}
\newcommand{\SJSci}{ $SIAM$ $J.$ $Scientific$ $Computing$}

\begin{document}

\title{A New Multidimensional Relativistic Hydrodynamics code based on 
Semidiscrete Central and WENO schemes.}

\author{Tanvir Rahman \altaffilmark{1}, Robert. B. Moore\altaffilmark{1}}

\altaffiltext{1}{Department of Physics, Rutherford Physics Building, 
McGill University, 3600 University Street, Montreal, QC H3A 2T8, Canada.}

\begin{abstract}
We have proposed a new High Resolution Shock Capturing (HRSC)
scheme for Special Relativistic
Hydrodynamics (SRHD) based on the semidiscrete central 
Godunov-type schemes and a modified 
Weighted Essentially Non-oscillatory (WENO) 
data reconstruction algorithm.   
This is the first application of the 
semidiscrete central schemes with high order 
WENO data reconstruction to the SRHD equations.  This 
method does not use a Riemann solver for flux computations and 
a number of one and two dimensional benchmark tests show that 
the algorithm is robust and comparable in 
accuracy to other SRHD codes.

\end{abstract}

%%%%%%%%%%%%%%%%%%%%%%%%%%%%%%%%%%%%%%%%%%%%%%%%%%%%%%%%
%%%%%%%%%%%%%%%%%%%%%%%%%%%%%%%%%%%%%%%%%%%%%%%%%%
\section{Introduction}
%%%%%%%%%%%%%%%%%%%%%%%%%%%%%%%%%%%%%%%%%%%%%%%%%%%%%%%
%%%%%%%%%%%%%%%%%%%%%%%%%%%%%%%%%%%%%%%%%%%%%%%%%%%

Gas flows at (ultra)relativistic speeds are an integral 
component of many astrophysical phenomena.  
Some significant examples of these 
include accretion disks around compact objects, 
Gamma ray bursts, collapse of (super)massive stars, 
pulsar wind nebula/supernova remnant interactions, 
Active Galactic Nuclei (AGN), X-ray binaries, 
superluminal jets and many others.  
As it stands, there is a massive amount of astronomical 
data from many of these sources that to date, have only been 
partially understood and which hold clues to many 
problems related to the phenomena mentioned above.  
Correlating these data with the theory 
in many of these phenomena would require doing simulations 
that would include modelling (ultra)relativistic gas/fluid 
flows.  Such flows are described by 
non-linear hyperbolic conservation laws that have 
shock waves as possible solutions.  
Shock waves are difficult to approximate numerically using 
standard finite difference techniques as they usually lead 
to spurious oscillations.  
High Resolution Shock Capturing (HRSC) schemes 
are a class of numerical methods devoted
specifically to approximating hyperbolic conservation laws.  
Over the past fifteen years HRSC research
have progressed immensely and they have been applied 
to a wide variety of problems involving classical 
and relativistic hydrodynamics.  For 
a pedagogical introduction to some modern 
HRSC schemes and their applications,  
see \cite{leveque2,toro,pen}.  To date,  the most commonly used 
HRSC schemes in relativistic hydrodynamics 
have been ones that have not incorporated many of the recent 
advances in HRSC research.  In this work, we will apply a
new scheme to the SRHD equations that is simpler than most of the 
previous approaches and incorporates a number of recent 
advances from HRSC research.  
In Rahman $\&$ Moore (2005), this scheme was applied to the 
multidimensional classical hydrodynamics code.  The work presented 
here should be considered an extension of that work.  
Before describing the motivations 
behind the algorithm used in this work, we provide 
a brief review of numerical methods used for 
the SRHD equations and that of HRSC schemes.        

Wilson J. R. (1972), Wilson et al. (1979), Hawley et al. (1984), 
and Centrella $\&$ Wilson (1984) were the first to apply 
numerical methods to approximate the SRHD equations.  
They discretized the SRHD equations using an 
explicit finite difference scheme with artificial 
viscosity and monotonic transport.  Since their work,  this 
technique has been used to 
study a number of astrophysical phenomena including stellar collapses, 
accretion disks, cosmology etc.  Most SRHD schemes up to 
the late eighties used the artificial viscosity 
technique of Wilson et al. (1979) to handle shock waves.  
However,  a major breakthrough occurred when 
HRSC schemes were applied to the SRHD equations 
(\cite{marti91,marquina92,eulderink93,eulderink95}).  
One of the most significant of these new schemes was the 
relativistic Piecewise Parabolic Method (PPM) 
by Mart\'i $\&$ M\"uller (1996) (MM) which was 
based on the PPM reconstruction scheme of 
Colella $\&$ Woodward  (1985).  For a recent review 
of modern SRHD schemes, we refer the reader to 
Mart\'i $\&$ M\"uller (2003).   

The use of HRSC schemes for SRHD equations essentially shifted the 
paradigm for solving these equations from artificial viscosity based 
methods to this approach.  Modern HRSC schemes are based on 
two conventional approaches for solving hyperbolic 
conservation laws.  These are 
the so-called upwind (\cite{Harten,vanleer,roe,toro}) 
and central methods (\cite{LxF,lax54,NT}).  
A common aspect of most modern HRSC-SRHD schemes is that 
they are solved using upwind schemes that use 
so-called Riemann solvers to approximate the solution.  A 
drawback of using Riemann solvers is that they are 
computationally expensive and difficult to implement 
because they need computations of eigenvalues and 
eigenvectors of the flux matrix and uses 
flux splittings etc., to compute the numerical fluxes.  
On the other hand, central schemes do not use Riemann solvers 
and are simpler and easier to implement.  
Until recently, central schemes were not 
considered for wide ranging applications because 
a trade-off for their simplicity was loss of accuracy.  
This is because central schemes were generally more dissipative than 
their upwind counterparts.  However, recent progress 
in HRSC research has addressed this issue and some newer 
formulations of central schemes have been shown 
to be comparable performance-wise to the upwind approach.  
The work presented here is based on 
some of these newer central schemes.  Before 
discussing them,  we provide below a brief overview of  
central schemes and discuss the most important aspect of 
HRSC schemes known as non-oscillatory 
data reconstructions methods.       
    
Modern central schemes are based on the first and 
second order shock capturing algorithms developed 
by Lax-Friedrichs (LxF) (\cite{LxF,lax54}), and 
Nessyahu $\&$ Tadmore (1990) (NT), respectively.  
Since NT, there have been a number of extensions of central
HRSC schemes to higher orders and multidimensions.  
Some of these recent extensions are summarized below.  
The NT scheme was extended to 3$^{rd}$ order by Liu $\&$ Tadmor (1998), 
and to multidimensions by Jiang $\&$ Tadmore (1998).  
These were followed by a new formulation of 
central HRSC schemes known as semidiscrete central schemes.  The 
semidiscrete schemes were designed to address the dissipation issue
mentioned above and were proposed in 
Kurganov $\&$ Tadmore (2000) (KT).  KT showed that 
semidiscrete schemes, which are 
formulated on non-staggered grids and use more 
accurate information of local speeds of propagation, 
are less dissipative than their staggered 
grid counterparts.  Hence, these schemes 
retained the simplicity of the central approach 
and were comparable in accuracy to other Riemann 
solver based upwind approaches.  The inception of 
semidiscrete central schemes  
precipitated a great deal research on 
extending this formalism which included higher order 
formulations in multidimensions,  genuinely multidimensional 
formulations and unstructured grid 
formulations among others(\cite{LT,GT,LPR1,LPR2,KP1,KNP,KL}).  

Besides the method used to advance the solution,  another 
aspect of HRSC schemes that is equally important for any algorithm 
is the so called non-oscillatory data interpolation method.  
Data interpolation is used in all HRSC schemes.  Non-oscillatory  
data reconstruction is the piecewise continuous 
polynomial interpolation of the 
data (which may contain discontinuities) over the 
computational domain.  Most of the HRSC schemes
mentioned above use the so called 
Essentially Non-Oscillatory (ENO) (\cite{HEOC}) 
data reconstruction algorithm for data interpolation.  
This methods works by interpolating the data using the smoothest 
stencil from a number of choices.  A modern extension of ENO schemes is 
the Weighted  Essentially Non-Oscillatory data 
reconstruction (WENO) algorithm (\cite{LOC,JS}) that has
a number of advantages over its predecessor.  In order 
to take advantage of the developments in central approaches and 
high order data reconstruction,  
another class of HRSC schemes were developed as an 
extension of the NT scheme by Levy et al. (\cite{LPR1,LPR2})  
called the Central Weighted Non-Oscillatory (CWENO) schemes.  
Recently, Kurganov $\&$ Levy (2000) (KL) have
combined the semidiscrete central schemes with the WENO 
reconstruction method and proposed yet another, better 
HRSC scheme.  The work presented here is 
based on the KL HRSC scheme which we will describe below.
 
Despite the many advances mentioned above,  
there have been relatively few attempts at 
applying central-type methods to 
computational astrophysics.  However this is rapidly changing.  Among the
most significant attempts to apply central schemes for computational
astrophysics, the following are noteworthy.  Del Zanna 
$\&$ Bucciantini (2002), have developed an algorithm that is 
a variation of the conventional finite volume central 
type approach and applied it to multidimensional 
SRHD and SRMHD equations.  They have also used these codes 
to study Pulsar Wind Nebula (PWN) interactions with the 
interstellar medium \cite{delzanna04} and supernova 
remnants \cite{delzanna03}.  Anninos $\&$ Fragile (2003,2004) have 
developed a new central type scheme for relativistic hydrodynamics 
in fixed, curved spacetime and have applied their scheme to study 
accretion disks around kerr black holes.  
Lucas-Serano et al. (2004) have assessed the 
applicability of central schemes to SRHD equations 
using the PPM data reconstruction scheme \cite{WC2}.  Their 
results were the first to demonstrate that central type 
schemes of KT can be used for SRHD.  
Recently, Shibata $\&$ Font (2005) have
successfully tested the suitability of central type schemes 
for general relativistic simulations.  We turn our 
attention now to applications of the WENO data 
reconstruction methods in computational astrophysics.  
Balsara (\cite{balsara}) have used the WENO methods 
extensively in magnetohydrodynamics.  Feng et al. (2004), 
have recently developed a cosmological hydrodynamics code 
using the WENO reconstruction scheme.  Zhang $\&$ MacFadyen (2005), 
have recently used the WENO scheme in their 
adaptive SRHD code.  The work mentioned above point to
a promising future for the application of central and WENO schemes
in astrophysical hydrodynamics.  Indeed, the work presented here was done
concurrently with many of the recent 
work mentioned above.  However, the algorithm 
presented here is the first to 
combine the WENO and the semidiscrete central 
approach for applications in SRHD.  We describe 
below this algorithm and its novel aspects.

In a previous paper, \cite{rahman1}, we 
have described in detail the incentive 
behind our algorithm for multidimensional 
classical hydrodynamics.  Many of the same reasons apply 
for the SRHD equations as well.  The algorithm is based on combining 
the semidiscrete approach of Kurganov $\&$ Levy (2000) 
with high order WENO data reconstruction methods.  To ensure 
robustness we have added the flattening, steepening and monotonicity 
preserving techniques of the PPM reconstruction 
scheme by Colella $\&$ Woodward (1985) to the 
data reconstruction scheme.  
Essentially, the simplicity of central schemes, the accuracy 
WENO reconstructions and the
robustness of the piecewise parabolic method 
have been combined to 
propose a new robust central scheme.  Building on 
the success of this algorithm for classical 
hydrodynamics,  it has been applied to the SRHD equations.  
This work is original in several respects.  It is the 
first attempt at solving the SRHD equations using 
the semidiscrete central scheme with
WENO data reconstructions.  The algorithm has been
tested using both 3$^{rd.}$ and 4$^{th.}$ order 
data reconstruction techniques.  
It is also a dimensionally unsplit algorithm and to our 
knowledge is the only unsplit multidimensional algorithm in SRHD.   
The work presented here can be considered as 
laying the foundation for the development of 
a multi-purpose, multi-dimensional SRHD code that could be used to
study a wide variety of astrophysical phenomena. 

The paper is organized as follows.  Sec. 2 presents the 
SRHD equations.  Sec. 3 describes the new algorithm for 
numerically approximating the SRHD equations.  In Sec. 4, we present
the results of a number of 
one dimensional benchmark tests.  In Sec. 5, the two 
dimensional tests are presented.  Finally we conclude 
this paper in Sec. 6.    

%%%%%%%%%%%%%%%%%%%%%%%%%%%%%%%%%%%%%%%%%%%%%%%%
%%%%%%%%%%%%%%%%%%%%%%%%%%%%%%%%%%%%%%%%%%%%%%%%%%%%%%%%%%
\section{Relativistic Hydrodynamic Equations}
\label{section:equations}
%%%%%%%%%%%%%%%%%%%%%%%%%%%%%%%%%%%%%%%%%%%%%%%%%%%%%%%
%%%%%%%%%%%%%%%%%%%%%%%%%%%%%%%%%%%%%%%%%%%%%%%%%%%
As a system of hyperbolic conservation laws, the equations of
special relativistic hydrodynamics are given by 
\begin{equation}
\frac {\partial {\bf U}({\bf w})} {\partial t} +
\frac {\partial {\bf f}^{i}({\bf w})} {\partial x^{i}} = 0 \quad .
\label{F}
\end{equation}
\noindent
where the indices run from 1 to 3. In the above equations 
${\bf U}$ and ${\bf f}^i$ are given by
\begin{eqnarray}
{\bf U}({\bf w})  &=& (D \quad , S^1 \quad , S^2 \quad , 
S^3 \quad , \tau) \quad , \\
 {\bf f}^{i}({\bf w}) & = &  (D v^{i} \quad ,
 S^j v^{i} + p \delta^{ij} \quad , S^i-Dv^i) \quad .
 \end{eqnarray}
Here $\delta^{ij}$ is the Kronecker delta, $v^i$ is the 
three velocity and $p$ is the pressure.  The relationship between the 
conserved variable and the primitive variable is given by
\begin{eqnarray}
D &=& \rho W \quad , \\
S^i &=& \rho h W^2 v^i \quad , \\
\tau &=& \rho h W^2 - p - D \quad ,
\end{eqnarray}
\noindent
where $h$ is the specific enthalpy; $h=1+\varepsilon+p/\rho$, and $W$ is
the Lorentz  factor satisfying 
$W \equiv u^0 = 1/\sqrt{1-v^2}$ with $v^2=v^iv_i$.  The 3-velocity 
components are obtained from the spatial 
components of the 4-velocity as $v^i={u^i}/{ u^0}$.  The equations above
are closed by the ideal gas Equation of State (EOS) given by,
\begin{eqnarray}
p= (\gamma-1) \rho \epsilon \quad ,
\end{eqnarray}
where $\gamma$ is the adiabatic index of the ideal gas and $\epsilon$ 
the energy density.  The difference
between the conserved and the primitive variables in these equations
require special treatment in the algorithm which will be discussed below.

%%%%%%%%%%%%%%%%%%%%%%%%%%%%%%%%%%%%%%%%%%%%%%%%%%%%%
%%%%%%%%%%%%%%%%%%%%%%%%%%%%%%%%%%%%%%%%%%%%%%%%%%%%%
\section{The Semidiscrete Central WENO Algorithm for SRHD}
\label{section:algorithm}
%%%%%%%%%%%%%%%%%%%%%%%%%%%%%%%%%%%%%%%%%%%%%%%%%%%%%%%%%%
%%%%%%%%%%%%%%%%%%%%%%%%%%%%%%%%%%%%%%%%%%%%%%%%
This section presents the new algorithm for 
solving the SRHD equations implemented in this work.  
This algorithm is very similar to that 
presented in a previous paper (\cite{rahman05}).  
Besides one major sub-step, that of recovering the 
primitive variables from the conserved variables, it is in fact
identical.  However, for the sake of completeness, a very brief
description of it is provided.  

As mentioned before, this new SRHD algorithm is based on the combination of 
the Semidiscrete  Central scheme of 
Kurganov $\&$ Levy (2000) (KL), and
the steepening, flattening and monotonicity preserving algorithm
of MM.  For a review and details of the derivation of KL, see 
\cite{KL}.  The reader is also referred to MM for 
details of the steepening, flattening and monotonicity preservation
algorithms.  In KL, the solution of a scalar hyperbolic conservation
law in one dimension,
\begin{eqnarray}
u_{t} + \frac {\partial f(u)} {\partial x}   = 0 \quad \nonumber \\ 
u(x,t=t^n)=u^n(x) \quad ,
\label{F1}
\end{eqnarray}
is given by
\begin{eqnarray}
\frac {d}{dt} \overline u_j (t) = - \frac {H_{i+\frac {1}{2}} (t) -
H_{i+\frac {1}{2}} (t) } {\Delta x} \quad ,
\label{updata}
\end{eqnarray}
where the flux $H_{i+\frac {1}{2}}$ is,
\begin{eqnarray}
H_{i+\frac {1}{2}} (t) := \frac { f(u^+_{i+\frac {1}{2}} (t)) + 
 f(u^-_{i+\frac {1}{2}} (t)) } {2} - 
\nonumber \\ \frac {a_{i+\frac {1}{2}} (t)} {2}
\left[ u^+_{i+\frac {1}{2}} - u^-_{i+\frac {1}{2}} \right] \quad .
\label{flux}
\end{eqnarray}
$u^{+}_{i+\frac {1}{2}}$, $a_{i+\frac {1}{2}}$ are
given by
\begin{eqnarray}
u^{-}_{i+\frac {1}{2}} = P_j^n (x_{j+1/2}) \quad ,
\end{eqnarray} 
\begin{eqnarray}
a_{i+\frac {1}{2}} := \max \{ \rho ( \frac {\partial f}{\partial u}
(u_{j+1/2}^{-}) ), \rho ( \frac {\partial f}{\partial u} 
(u_{j+1/2}^{+}) ) \} \quad ,
\end{eqnarray}
where $\{P_j^n(x)\}$ is a non-oscillatory piecewise polynomial
reconstruction of cell averaged data at time $t=t^n$, 
$\{\overline u_j^n(x) \}$.  In this paper, both 
3$^{rd.}$ and 4$^{th}$ order WENO reconstruction
methods of \cite{KL,LPR1} are used.  
In Eq. 10,  $a_{i+\frac {1}{2}}$ is the speed of propagation of
$u$ at the interface of a cell that is determined from 
the spectral radius of the Jacobian of the flux $f$.  For a 
system of equations 
in multidimensions, the scheme given above can be extended as follows.
A dimensionally unsplit approach can be used to extend Eq. 9 to 
multidimensions.  In this approach,  Eq. 9 is modified by 
adding another term corresponding to the 
flux differences in the y-direction.  For a system of equations, the 
scalar $f$ is replaced by a vector ${\bf F}$ and 
$\frac {\partial f}{\partial u}$ is replaced by the Jacobian of 
${\bf F}$.  

For all the computations presented here, the 
piecewise polynomial interpolations 
have been done on the primitive variables from which the conserved
variables are computed using Eqns. 4-6.  In each step of 
our integration routine,  once the conserved 
variables are updated, the primitive variables 
are immediately recovered.  
The method used to do this is by solving of a non-linear equation 
using the Newton-Raphson root finding method 
(see \cite{marti03} for details).  
It is this aspect of numerical SRHD that makes 
them computationally more expensive than
the Euler equations of gas dynamics.  We can now 
summarize our algorithm
in the following few steps:

{\bf step 1}: \textit {Given initial data 
${\bf {\overline u}_{j}^n}$ (primitive variables), 
use an $nth.$ (for n $\le$ 4) order WENO
reconstruction algorithm to 
construct a piecewise polynomial interpolation
of each variable.} 

{\bf step 2}: \textit{Apply the steepening (only to the density 
$\rho$), flattening and monotonicity
preserving algorithms to the interface 
values of the primitive variables.} 

{\bf step 3}: \textit{Update ${\overline u}_{j}^n$ (the conserved
variables) to  ${\overline u}_{j}^{n+1}$ using Eqns. 9 and 10 above.}

{\bf step 4}: \textit{Recover the primitive variables 
from the updated conserved variables.} 

{\bf step 5}: \textit{Repeat steps 1 to step 4}

The time integration of Eq. 9 is done using 
a high order total variation diminishing (TVD) Runge-Kutta scheme 
\cite{SO}, which combines the first order forward Euler 
method with predictor-corrector steps.  This is,
\begin {eqnarray}
U^{(1)} = U^n + \Delta T L (U^n) \nonumber \\
U^{(2)} = \frac {3}{4} U^n + \frac {1}{4}U^{(1)} + \frac {3}{4} \Delta t
L(U^{(1)}) \nonumber \\
U^{n+1} = \frac {1}{3} U^n + \frac {2}{3} U^{(2)} +  \frac {2}{3} \Delta
t L(U^{(2)}) \quad . 
\end{eqnarray}

%%%%%%%%%%%%%%%%%%%%%%%%%%%%%%%%%%%%%%%%%%%%%%%%%%%
%%%%%%%%%%%%%%%%%%%%%%%%%%%%%%%%%%%%%%%%%%%%%%%%%%%%%%%
\section{Numerical Tests}
\label{section:results1D}
%%%%%%%%%%%%%%%%%%%%%%%%%%%%%%%%%%%%%%%%%%%%%%%%%%%%%%%
%%%%%%%%%%%%%%%%%%%%%%%%%%%%%%%%%%%%%%%%%%%%%%%%%%%

SRHD schemes
are usually tested by a series of standard benchmark problems.  The most
important of these are the so-called $shock$ $tube$ Riemann
problems.  A shock tube test in one dimension can be described as follows; 
a one dimensional pipe is divided into two halves by a membrane and the
thermodynamic states of each half is specified.  When the 
membrane between the halves is removed, 
depending on the initial conditions, contact discontinuities, 
shocks and rarefaction waves etc. will result.  
Analytic solutions of the amplitude and position 
of these features have been derived 
for both classical and relativistic
hydrodynamics.  For analytical solutions of the relativistic 
hydrodynamics shock tube problems, see \cite{marti94,thompson86,pons00}.  
For a list of some standard shock tube tests 
along a review of the performance of some SRHD schemes, see 
Mart\'i $\&$ M\"uller (2003).  
In addition to the Riemann problems, we have done some 
other tests that do not have analytic solutions.  In each case,
the results are compared to the literature.  

In the following two sections, we present the results of 
the benchmark tests in one and two dimensions.  
Before we present our results, we list 
in Table 1, the input parameters we have
used in the tests for contact steepening, flattening and the monotonicity 
preserving steps.  We have set $\gamma = 5/3$ in all 
the computations, unless otherwise specified.

%%%%%%%%%%%%%%%%%%%%%%%%%%%%%%%%%%%%%%%%%%%%%%%%%%%%%%%%%%
%%%%%%%%%%%%%%%%%%%%%%%%%%%%%%%%%%%%%%%%%%%%%%%%
\subsection{One Dimensional tests}
%%%%%%%%%%%%%%%%%%%%%%%%%%%%%%%%%%%%%%%%%%%%%%%%%%%%%%%%%%%%%
%%%%%%%%%%%%%%%%%%%%%%%%%%%%%%%%%%%%%%%%%%%%%

{\bf Problem 1:}

The initial states of the right and left halves of the domain are
given by $p_{\rm L}=1.0,  \rho_{\rm L}=1.0,  v_{\rm L}=0.9$ (left) 
and $p_{\rm R}=10.0,  \rho_{\rm R}=1.0,  v_{\rm R}=0$ (right).  This
initial condition leads to a strong reverse shock.
In Figure 1,  the results are shown for both 3$^{rd}$  and 4$^{th}$  
order reconstructions.  We have shown the density,
velocity and pressure profiles at t=0.4 and compared the 
numerical results to analytic solutions.  Excellent agreement
between these results and the analytic solutions is noted.  Also, these 
results match closely with those obtained by Lucas-Serano
et al. (2004) which were obtained using the PPM.  
As was the case in Lucas-Serano et al. (2004), small oscillations 
noted behind the shocks disappear completely 
as the CFL number is lowered.  However, such oscillations 
remain more pronounced for the 4$^{th}$  order reconstruction.  
In Table 2,  $L^1$ errors for the density for 
both 3$^{rd}$ and 4$^{th}$ order 
reconstructions are shown.  It shows that the order of the 
scheme is approximately one, as is 
expected for solutions with shocks.

%%%%%%%%%%%%%%%%%%%%%%%%%%%%%%%%%%%%%%%%%%%%%%%%%%%%%%%%%%%%%
%%%%%%%%%%%%%%%%%%%%%%%%%%%%%%%%%%%%%%%%%%%%%
{\bf Problem 2:}
%%%%%%%%%%%%%%%%%%%%%%%%%%%%%%%%%%%%%%%%%%%%%%%%%%%%%%%%%%%
%%%%%%%%%%%%%%%%%%%%%%%%%%%%%%%%%%%%%%%%%%%%%%%

The initial states of the right and left halves of the domain are 
$p_{\rm L}=10.0, \rho_{\rm L}=1.0, v_{\rm L}=-0.6$ (left) and
$p_{\rm R}=10.0, \rho_{\rm R}=1.0, v_{\rm R}=0.5$ (right).  In
Figure 2, the results using 
the 3$^{rd}$ and 4$^{th}$ order WENO reconstructions are presented.  
Once again, we note that the shock positions
and velocities are captured well by both reconstruction schemes.  Unlike
the previous test, post shock oscillations are not seen in either of the
reconstructions.  Direct comparison with Lucas-Serano et al. (2004), 
show good agreements as well.  
Table 3 presents the $L^1$ errors of the density for both 3$^{rd}$
and 4$^{th}$ order reconstructions.  The order is 
approximately one for both cases except for very high grid spacings.

%%%%%%%%%%%%%%%%%%%%%%%%%%%%%%%%%%%%%%%%%%%%%%%%%%%%%%%
%%%%%%%%%%%%%%%%%%%%%%%%%%%%%%%%%%%%%%%%%%%%%%%%%%%
{\bf Problem 3:}
%%%%%%%%%%%%%%%%%%%%%%%%%%%%%%%%%%%%%%%%%%%%%%%%%%%%%%%%%
%%%%%%%%%%%%%%%%%%%%%%%%%%%%%%%%%%%%%%%%%%%%%%%%%

The initial states of the right and left handed halves of the domain are
given by $p_{\rm L}=13.3,  \rho_{\rm L}=10.0,  v_{\rm L}=0$ (left) and
$p_{\rm R}=0,  \rho_{\rm R}=1.0,  v_{\rm R}=0$ (right).  We expect a 
shock wave, a contact discontinuity and a rarefaction wave from this
test.  Figure 3 shows the results of the computation.   
There is good agreement between the computed and analytic 
solutions.  Direct comparison with Lucas-Serano et al. (2004) and 
Zhang $\&$ MacFadyen (2005) show good agreements as well.  In Table 4, 
the $L^1$ errors of the density for both for 3$^{rd}$
and 4$^{th}$  order reconstructions are shown.  Although irregular, 
these values show similar trends and magnitudes to those obtained 
by others (see Lucas-Serano
et al. (2004), Zhang $\&$ MacFadyen (2005)).  We believe the error could be
optimized by appropriately fine tuning the 
smoothing and flattening parameters.

%%%%%%%%%%%%%%%%%%%%%%%%%%%%%%%%%%%%%%%%%%%%%%%%%%%%%%%%%%
%%%%%%%%%%%%%%%%%%%%%%%%%%%%%%%%%%%%%%%%%%%%%%%%
{\bf Problem 4:}
%%%%%%%%%%%%%%%%%%%%%%%%%%%%%%%%%%%%%%%%%%%%%%%%%%%%%%%
%%%%%%%%%%%%%%%%%%%%%%%%%%%%%%%%%%%%%%%%%%%%%%%%%%%

The initial states of the right and left handed halves of the domain are
given by $p_{\rm L}=1000.0,  \rho_{\rm L}=1.0,  v_{\rm L}=0$ (left) 
and $p_{\rm R}=0.01,  \rho_{\rm R}=1.0,  v_{\rm R}=0$ (right).  The initial
condition gives rise to a right-moving shock, a left-moving rarefaction
wave and a contact discontinuity in between.  This
is a fairly demanding test due to the $10^5$ order pressure ratio.
In Figure 4, we show the results of our computation.  As expected the 
3$^{rd}$  order scheme is more 
dissipative than the 4$^{th}$  order scheme.  Once
again, a direct comparison between the results and 
analytic solutions show comparable agreements to 
the  same tests done with other schemes (see Lucas-Serano
et al. (2004), MacFadyen and Zhang (2005)).  
The  $L^1$ errors (Table 5) of the density also 
show similar results.  

%%%%%%%%%%%%%%%%%%%%%%%%%%%%%%%%%%%%%%%%%%%%%%%%%%%%%%%%
%%%%%%%%%%%%%%%%%%%%%%%%%%%%%%%%%%%%%%%%%%%%%%%%%%
{\bf Problem 5 (Shock Reflection Test)}
%%%%%%%%%%%%%%%%%%%%%%%%%%%%%%%%%%%%%%%%%%%%%%%%%%%%%%%%%%%
%%%%%%%%%%%%%%%%%%%%%%%%%%%%%%%%%%%%%%%%%%%%%%%
We consider now an even more severe test for the scheme.  
The shock reflection test consists of an ultra-cold 
relativistic wind hitting a solid wall and a reflected 
shock wave propagating backward leaving a static region 
of hot gas.  The Lorenz factor of the wind is set by the 
precision level of the code.  Hence, winds at very high 
speeds are considered.  To directly compare our results 
with that of Del Zanna $\&$ Bucciantini (2002), 
Lucas-Serano et al. (2004), and Zhang $\&$ MacFadyen (2005), 
we have chosen the following initial conditions; $p=.01,  \rho=1.0,
 v=0.99999,  v=0.9999999999$ for the first and second tests, respectively.
The solid wall is placed at the right edge of the 
computational domain.  The post shock density is an increasing function
of the initial flow velocity.  The compression ratio $\sigma 
= \rho_2/\rho_1$, satisfies 
\begin{equation}
\sigma= \frac {\gamma + 1}{\gamma - 1} + \frac {\gamma}{\gamma - 1} \epsilon_2
\end{equation}
where, $\epsilon_2 = W_1-1$ and  $W_1$ is the initial Lorentz factor.  We
have used $\gamma= 4/3$ for this test.
The results are shown in 
Figures 5 and 6.  Consider first Figure 5, which corresponds to a
Lorenz boost factor, $W = 224$.  Direct comparison between our results
and those of Lucas-Serano et al. (2004), and Del Zanna $\&$ Bucciantini
(2002), show good qualitative agreement.  Note that both 3$^{rd}$  and
4$^{th}$  order reconstruction schemes capture the shock wave without
oscillations, in contrast to the case of Del Zanna and Bucciantini (2002), 
where the 3$^{rd}$  order CENO scheme fails.  Next,  consider 
the case of $W = 70710.675$ (Figure 6).  Comparing our results to those
of Zhang $\&$ MacFadyen (2005), there is excellent qualitative agreement.
These results show that the scheme is robust against 
both 3$^{rd}$  and 4$^{th}$ order reconstructions for 
ultra-relativistic flows.

%%%%%%%%%%%%%%%%%%%%%%%%%%%%%%%%%%%%%%%%%%%%%%%%%%%%%%
%%%%%%%%%%%%%%%%%%%%%%%%%%%%%%%%%%%%%%%%%%%%%%%%%%%%
{\bf Problem 6: Mixed Solution test}
%%%%%%%%%%%%%%%%%%%%%%%%%%%%%%%%%%%%%%%%%%%%%%%%%%%%%%%%%%%%%%
%%%%%%%%%%%%%%%%%%%%%%%%%%%%%%%%%%%%%%%%%%%%

For the final one dimensional test, we followed 
Del Zanna and Bucciantini (2002), and considered a test 
in which oscillatory and smooth 
solutions arise close together simultaneously.  
The initial configuration is given by;
\begin{eqnarray}
p_{L}=50.0, \rho_{L}=5.0, \qquad v_{L}=0  (\rm{left}) \quad , \\
p_{R}=5.0, \rho_{R}= 2.0 + 0.3 \sin 50x, \qquad v_{R}=0  (\rm{right})  \quad .
\end{eqnarray}
The solution consists of the interaction between a 
blast wave and a density wave.  The results are shown in Figure 7, which 
shows the density profile for both the 3$^{rd}$  and 4$^{th}$  order
reconstructions simultaneously.  We see that the 4$^{th}$  order scheme gives
more oscillatory results than 3$^{rd}$  order WENO.  Direct comparison with
Del Zanna $\&$ Bucciantini (2002),  show good agreement.

%%%%%%%%%%%%%%%%%%%%%%%%%%%%%%%%%%%%%%%%%%%%%%%%%%%%%%%%%%%%%%%%%%%%%%%%%%%%
%%%%%%%%%%%%%%%%%%%%%%%%%%%%%%
\section{2-dimensional Relativistic Hydrodynamics}
\label{section:results2D}
%%%%%%%%%%%%%%%%%%%%%%%%%%%%%%%%%%%%%%%%%%%%%%%%%%%%%%%%%%%%%%%%%%%%%
%%%%%%%%%%%%%%%%%%%%%%%%%%%%%%%%%%%%%

The one dimensional scheme is extended to two 
dimensions in a straightforward manner using a 
dimensionally unsplit scheme for advancing 
the solution that was described earlier.  The 
two dimensional reconstruction is done 
using dimensional splitting.  
For the multidimensional tests, we consider the 
a two dimensional Riemann problem, a symmetric blast wave and a
relativistic jet.  All of these tests are compared 
to previous results.

%%%%%%%%%%%%%%%%%%%%%%%%%%%%%%%%%%%%%%%%%%%%%%%%%%%%%%%%%%%%%%%%%%%%%%%
%%%%%%%%%%%%%%%%%%%%%%%%%%%%%%%%%%%
\subsection{Two-dimensional shock tube test}
%%%%%%%%%%%%%%%%%%%%%%%%%%%%%%%%%%%%%%%%%%%%%%%%%%%%%%%%%%%%%%%%%%
%%%%%%%%%%%%%%%%%%%%%%%%%%%%%%%%%%%%%%%%

Following Del Zanna $\&$ Bucciantini (2002), Lucas-Serano et al. (2004), 
and Zhang and MacFadyen (2005),  consider the following 
two-dimensional Riemann Problem.  A two dimensional square 
region is divided into four quadrants and the initial 
conditions given by
\begin{eqnarray}
(\rho,v_x,v_y,p)^{NE}=&(0.1,0.0,0.0,0.01) \quad ,\nonumber \\
(\rho,v_x,v_y,p)^{NW}=&(0.1,0.99,0.0,1.0) \quad ,\nonumber \\
(\rho,v_x,v_y,p)^{SW}=&(0.5,0.0,0.0,1.0)\quad , \nonumber \\
(\rho,v_x,v_y,p)^{SE}=&(0.1,0.0,0.99,1.0)\quad .
\end{eqnarray}
Outflow boundary conditions are used throughout the domain and 
the adiabatic gas index is set to $\gamma$ = 5/3.  The resulting 
pattern is shown in Figure 8 which were computed 
using the 4$^{th}$  order WENO and 
PPM (from Lucas-Serano et al. (2004)) reconstruction schemes. 
The solutions are shown at time time $t=0.4$.  The PPM based 
calculations have been
presented here for the sake of direct comparison.  
The salient features of these results are as follows.
Note that the WENO based scheme gives 
sharper contact discontinuities than the PPM scheme 
for the same grid resolution.  Around the head of the shock,   
we have been able to resolve both the bow 
shock and the inner shock structure.  Around the tails, 
where the shock merge, we are able resolve the jet-like 
structure seen due to the merging of the two discontinuities.

%%%%%%%%%%%%%%%%%%%%%%%%%%%%%%%%%%%%%%%%%%%%%%%%%%%%%
%%%%%%%%%%%%%%%%%%%%%%%%%%%%%%%%%%%%%%%%%%%%%%%%%%%%%
\subsection{Spherically symmetric blast-wave test}
%%%%%%%%%%%%%%%%%%%%%%%%%%%%%%%%%%%%%%%%%%%%%%%%%%%%%%%
%%%%%%%%%%%%%%%%%%%%%%%%%%%%%%%%%%%%%%%%%%%%%%%%%%%
In this test,  a cylindrically symmetric blast wave is studied.  
The set-up of the problem is as follows.  The domain 
is $[-1,1] \times [-1,1]$ with outflow boundary conditions.  
The initial configuration is given by
\begin{eqnarray}
(\rho,v_x,v_y,p)=(1.,0.,0.,1000.) \quad {\rm if} \qquad 
r^2 \quad < \quad (.4)^2 \nonumber \\
(\rho,v_x,v_y,p)=(1.,0.,0.,1.0) \quad {\rm Otherwise} \quad .
\end{eqnarray}
This problem does not have an analytic solution so the results are
compared to a one dimensional simulation.  The results are also compared 
to Del Zanna $\&$ Bucciantini (2002), and Zhang $\&$
MacFadyen (2005).  The initial condition given above give rise to a
cylindrically symmetric blast wave moving outward, a rarefaction wave
moving inward and and a contact discontinuity.  A 4$^{th}$  order 
WENO data reconstruction algorithm was used for this test.  
In Fig. 9,  the test results are shown at time $t=0.4$.  Direct 
comparison with the above mentioned references 
show excellent agreement.  Cylindrical symmetry is
also well preserved in the computations.

%%%%%%%%%%%%%%%%%%%%%%%%%%%%%%%%%%%%%%%%%%%%%%%%%%%%%%%%%%%%%%%%%%%%%%%%%%%%%%%%%%%%%%%%%%%%%%%%%%%%%%%%%%
\subsection{2D Relativistic Jet}
%%%%%%%%%%%%%%%%%%%%%%%%%%%%%%%%%%%%%%%%%%%%%%%%%%%%%%%%%%%%%%%%%%%%%%%%%%%%%%%%%%%%%%%%%%%%%%%%%%%%%%%%%%
The final multidimensional test undertaken for our algorithm is
that of relativistic jets in planar geometry.  The dynamics 
and morphology of
jets have been studied by many authors 
(see Zhang $\&$ MacFadyen (2005), and references therein).  This is 
a standard benchmark for testing any new relativistic HRSC scheme.  
To test the ability of our codes to capture the salient features of
relativistic jets, we have considered the following scenario.  Our tests 
are done in planar geometry.  We consider a rectangular domain of length
$[0,45] \times [-25,25]$.  Through a small opening along the axis of
symmetry we inject a relativistic fluid at $v_x=0.99c, V_y=0.0$.  The
pressure between the inflow and the ambient medium is considered to be
at equilibrium.  And we take a density ratio of the inflow and ambient
medium of $0.01$.  Reflective boundary conditions
are used along the axis-of symmetry.  Otherwise outflow boundary
conditions are used elsewhere.   

Figs. 10-11 progression of the jets at times $t=5$, $t=25$, $t=50$ 
and $t=75$ respectively.  Comparing our results to Lucas-Serano (2004),
Del Zanna $\&$ Bucciantini (2002), and Zhang $\&$ MacFadyen (2005), we
find good qualitative agreement.  The jets capture the expected
morphology quite well.  There is a supersonic jet that extends from the
nozzle and impacts with the ambient medium, at which point we note a
terminal planar shock, a contact discontinuity and a bow shock.  There is 
a cocoon that separates the ambient medium from the jet 
material by a contact discontinuity where we note 
Kelvin-Helmholtz instabilities.  We also note the absence of
$carbunckles$, that is purely a numerical artifact.          

Figs. 12-13, show snapshots of the same calculations using 
a slightly narrower opening of the in flowing jets.  Compared to the 
previous case, the jets are more collimated in this case. 

%%%%%%%%%%%%%%%%%%%%%%%%%%%%%%%%%%%%%%%%%%%%%%%%%%%%%%%%%%
%%%%%%%%%%%%%%%%%%%%%%%%%%%%%%%%%%%%%%%%%%%%%%%%
\section{Summary and Conclusion}
\label{section:summary}
%%%%%%%%%%%%%%%%%%%%%%%%%%%%%%%%%%%%%%%%%%%%%%%%%%%%
%%%%%%%%%%%%%%%%%%%%%%%%%%%%%%%%%%%%%%%%%%%%%%%%%%%%%%

We have developed a new multidimensional relativistic hydrodynamics code
based on the semidiscrete central and WENO 
reconstruction approach, and some elements of the PPM method.  This 
is the first such multidimensional 
relativistic code combining these elements.  
A dimensionally unsplit scheme was used to advance the 
solution.  We also carried
out a number of one and two dimensional tests of the code using both 
3$^{rd}$ and 4$^{th}$ order reconstruction methods.  The test 
results are comparable to a number of other codes currently being used
in computational astrophysics.  In all cases,  the tests showed 
good agreements with codes currently in use.  

The work presented here should be considered as the first few steps 
towards the development of a multi-purpose relativistic hydrodynamics
code.  Some of the developments 
mentioned above on semidiscrete central schemes could well extend the
performance of our code.  These include the 
genuinely multidimensional formulation,  the formulation 
on unstructured grid or higher order WENO reconstructions.  
From the point of applications,  we 
are working to extend our code to three dimensions, provide it with
adaptive capabilities and apply it for parallel architecture using MPI.

\begin{acknowledgements}

The authors would like to thank Martin Gander for reviewing 
and advising during various stages of this project 
despite extenuating circumstances.  
T.R. would like to thank Jose Font,  Andrew MacFadyen 
and Chris Fragile for some useful discussions.  T.R. would also 
also like to thank Gil Holder for his interest in the work.  
T.R. would like to thank Steve Liebling of C.W. Post campus 
of Long Island University for hospitality during part of this work.  
This work was supported by the National Sciences 
and Engineering Research Council (NSERC) of Canada.

\end{acknowledgements}

\newpage

\begin{center}
\begin{table}[!h]
\centering

\begin{tabular}{cccccccc}
\hline
Test & $K_0$ & $\eta^{(1)}$ & $\eta^{(2)}$ & $\epsilon^{(1)}$ & $\omega^{(1)}$ & 
$\omega^{(2)}$ & $\epsilon^{(2)}$ \\
 \hline
Test 1 &  1.0 & 5.0  & 0.05 & 0.7 & 0.52 & 10.0 & 5.  \\
Test 2 &  1.0 & 5.0 & 0.05 & 0.7 & 0.52 & 10.0 & 5.0 \\
Test 3 &  1.0 & 5.0 & 0.05 & 0.7 & 0.52 & 10.0 & 5.0 \\
Test 4 &  1.0 & 5.0  & 0.05 & 0.7 & 0.52 & 10.0 & 5.0 \\
Test 5 &  1.0 & 5.0  & 0.1  & 0.1 & 0.52 & 10.0 & 0.0001 \\
\hline
\end{tabular}

\label{tabla:ppm_param}
\caption{Values of the flattening, steepening and monotonicity
preserving parameters used in 
one dimensional shock tube tests presented in Sec. 5.1}
\end{table}
\end{center}

\begin{center}
\begin{table}[!h]
\centering

\begin{tabular}{|c|cccc|}
\hline
N & $L^1$-error (4$^{th}$ order) & Rate & 
$L^1$-error (3$^{rd}$ order) & Rate \\
\hline
40 & 2.82E-1 & - & 3.17E-1  & - \\
80 & 1.33E-1 & 1.08  & 1.55E-1 & 1.06   \\
160 & 6.77E-2 & .9466  & 8.04E-2  & 0.96 \\
320 & 3.4E-2 & 1.02 & 4.22E-2  & 0.94 \\
640 & 1.75E-2 & 0.98 & 2.32E-2 & 0.92 \\
\hline
\end{tabular}

\caption{$L^1$ errors of density in Test 1, Sec. 5.1. Shown are the 
errors in density for both 3$^{rd}$ and 4$^{th}$ order reconstruction.}
\end{table}
\end{center} 

\begin{center}
\begin{table}[!h]
\centering

\begin{tabular}{|c|cccc|}
\hline
N & $L^1$-error (4$^{th}$ order) & Rate & $L^1$-error (3$^{rd}$ order) & Rate \\
\hline
40 & 2.53E-1  & - & 3.76E-1 & - \\
80 & 1.66E-1 & 0.63  & 2.51E-1 & 0.61  \\
160 & 7.31E-2 & 1.22  & 1.26E-1 & 0.99  \\
320 & 3.77E-2 & 0.98 & 6.55E-2  & 0.95 \\
640 & 1.92E-2 & 1.15  & 3.33E-2 & 0.99 \\
\hline
%\label{table2}
%\caption{L^1 and L^{\infty} errors for 4$^{th}$ order reconstruction}
\end{tabular}

\caption{$L^1$ errors for Test 2, Sec. 5.1. Shown are the 
errors in density for both 3$^{rd}$ and 4$^{th}$ order reconstruction.}
\end{table}
\end{center}

\begin{center}
\begin{table}[!h]
\centering 

\begin{tabular}{|c|cccc|}
\hline
N & $L^1$-error (4$^{th}$ order) & Rate & $L^1$-error (3$^{rd}$ order) & Rate \\
\hline
40 & 4.40E-1  & - & 4.98E-1 & - \\
80 & 2.28E-1 & 0.95  & 2.80E-1 & 0.84   \\
160 & 1.30E-1 & 0.81 & 1.49E-1  &  0.92  \\
320 & 6.19E-2 & 1.07 & 7.23E-2  &  1.04 \\
640 & 3.23E-2 & 0.94 & 3.77E-2 & 0.94 \\
\hline
\end{tabular}
\caption{$L^1$ errors for Test 3, Sec. 5.1. Shown are the 
errors in density for both 3$^{rd}$ and 4$^{th}$ order reconstruction.}
\end{table}
\end{center} 

\begin{center}
\begin{table}[!h]
\centering

\begin{tabular}{|c|cccc|}
\hline
N & $L^1$-error (4$^{th}$ order) & Rate & $L^1$-error (3$^{rd}$ order) & Rate \\
\hline
80 & 2.79E-1 & -  & 2.81E-1  &  - \\
160 & 1.85E-1 & 0.51 & 1.85E-1  & 0.6   \\
320 & 1.51E-1 & 0.29 & 1.50E-1  & 0.3  \\
640 & 1.01E-1 & 0.57 & 9.42E-2 & 0.66 \\
\hline
\end{tabular}
\caption{$L^1$ errors for Test 3, Sec. 5.1. Shown are the 
errors in density for both 3$^{rd}$ and 4$^{th}$ order reconstruction.}
\end{table}
\end{center} 

\begin{figure}[!h]
  \begin{center}
  \includegraphics[angle=-90,width=0.8\textwidth]{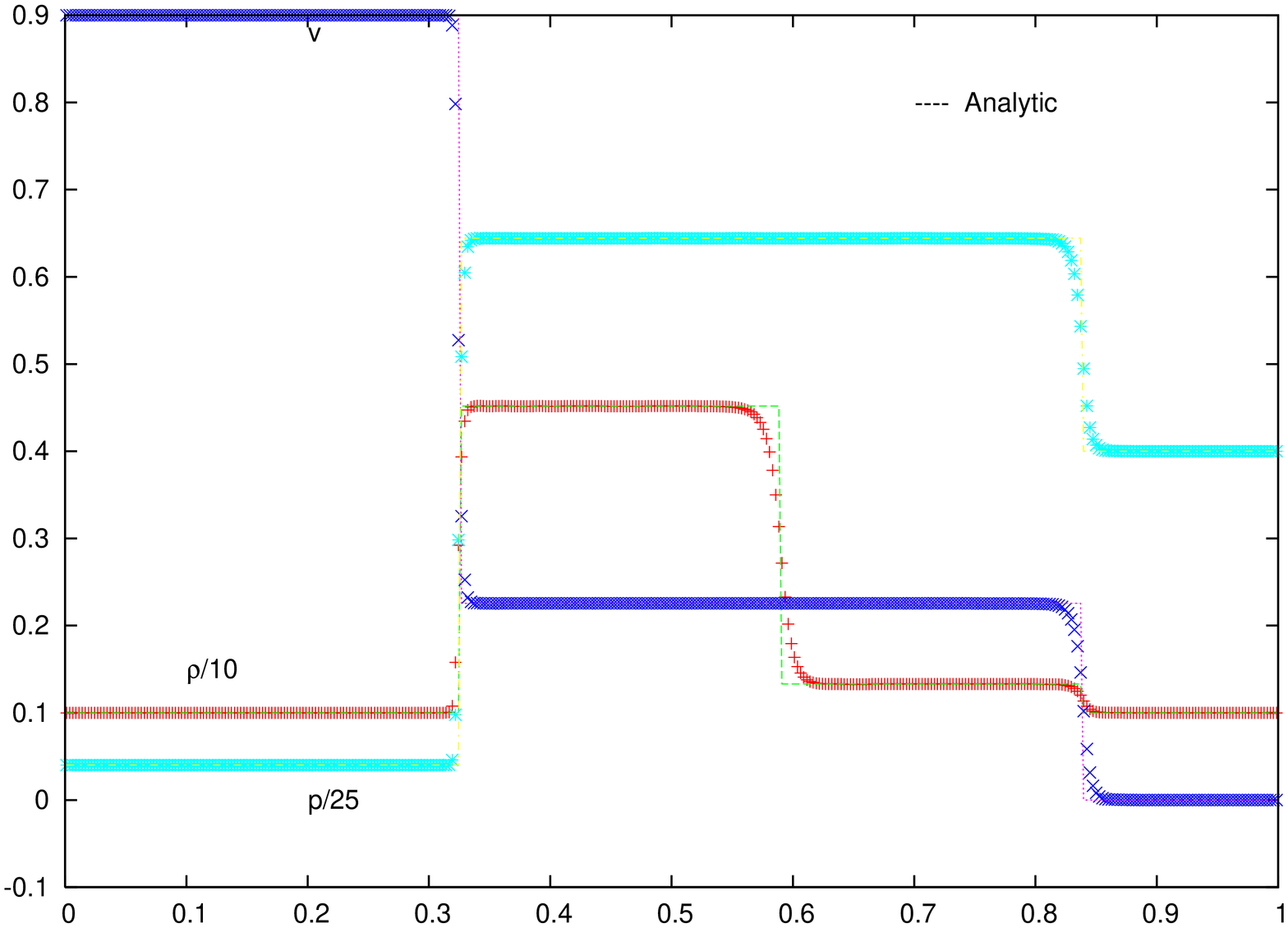}\\
  \includegraphics[angle=-90,width=0.8\textwidth]{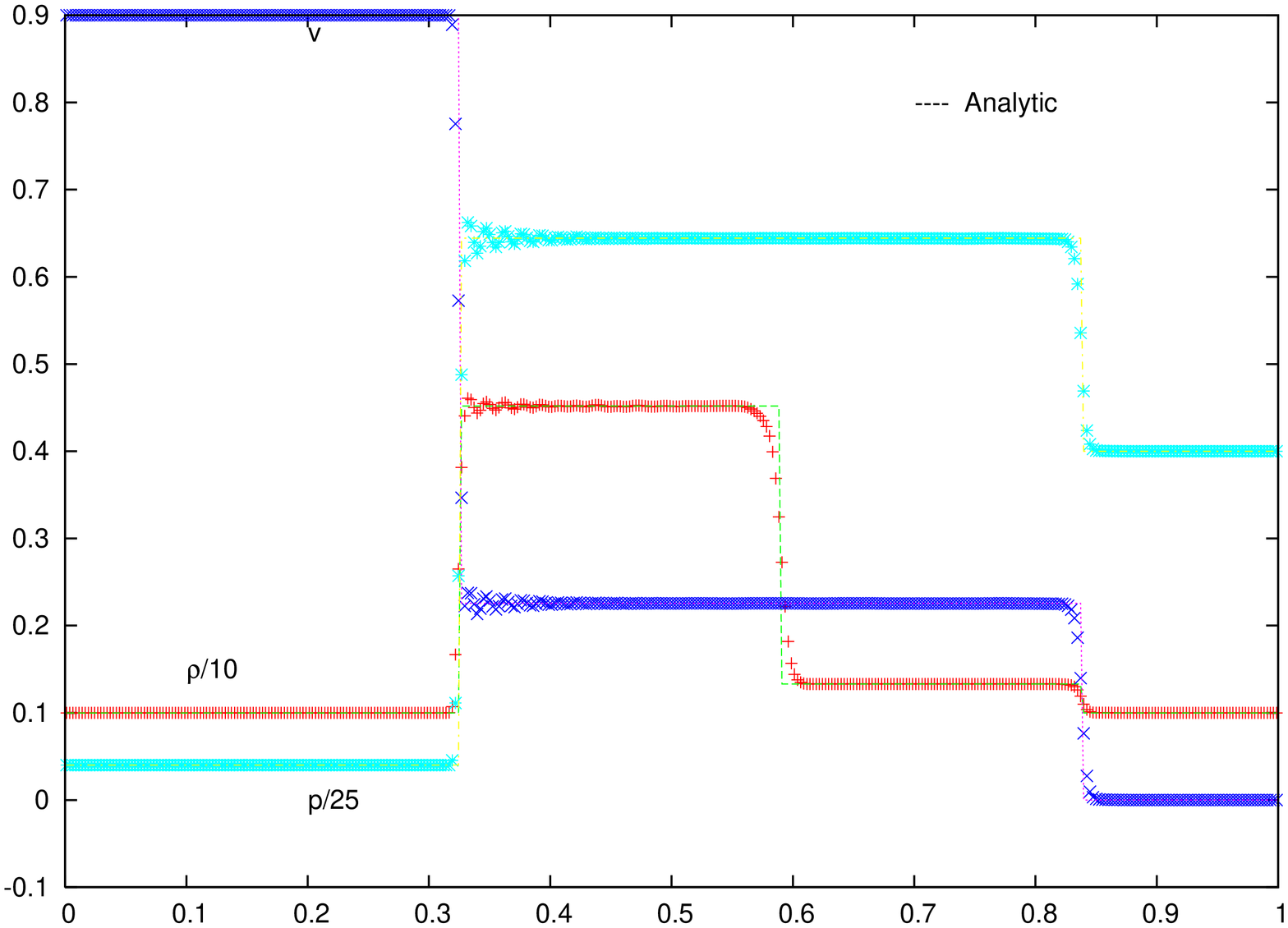}
  \caption{Density, Pressure and velocity profiles 
for test 1 (Sec. 5.1) at  t=0.4 using
3$^{rd}$ (top) and 4$^{th}$ (bottom) order WENO reconstructions and n=400.}
  \label{figs1_2}
  \end{center}
\end{figure}

\begin{figure}[!h]
  \begin{center}
  \includegraphics[angle=-90,width=0.8\textwidth]{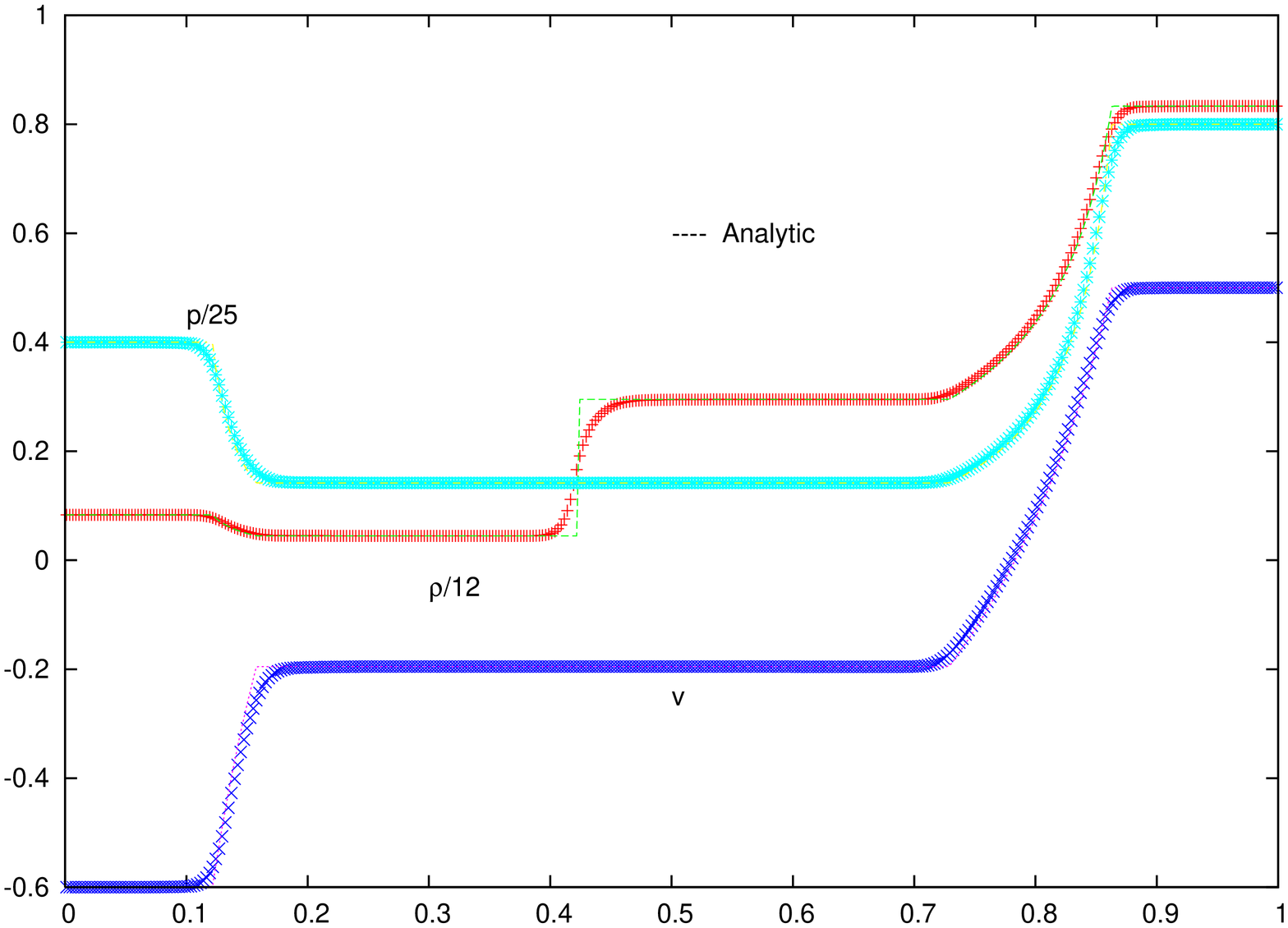}\\
  \includegraphics[angle=-90,width=0.8\textwidth]{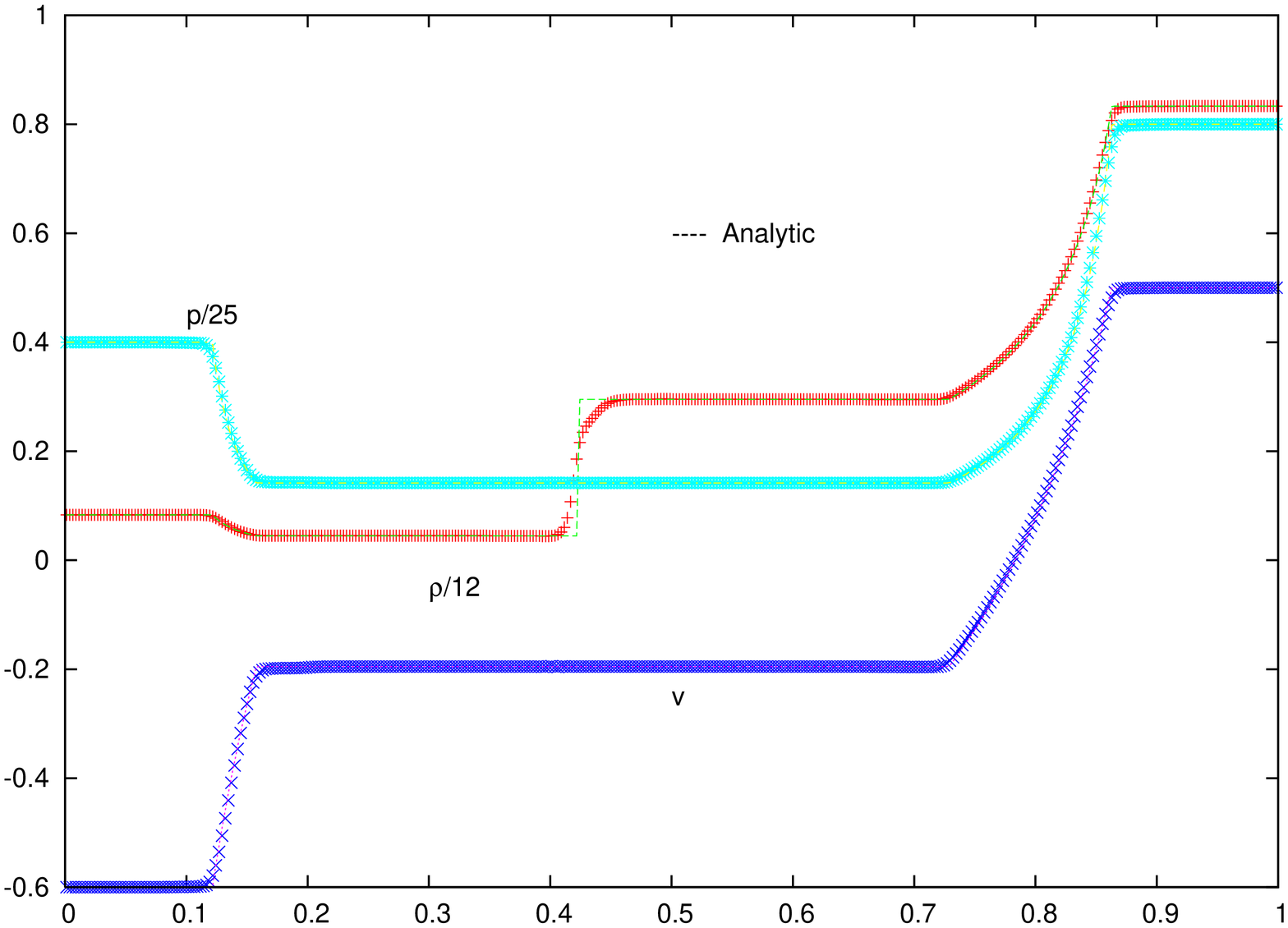}
  \caption{Density, Pressure and velocity profiles for 
test 2 (Sec. 5.1) at t=0.4 using
3$^{rd}$ (top) and 4$^{th}$ (bottom) order WENO reconstructions and n=400.}
  \label{figs3_4}
  \end{center}
\end{figure}

\begin{figure}[!h]
  \begin{center}
  \includegraphics[angle=-90,width=0.8\textwidth]{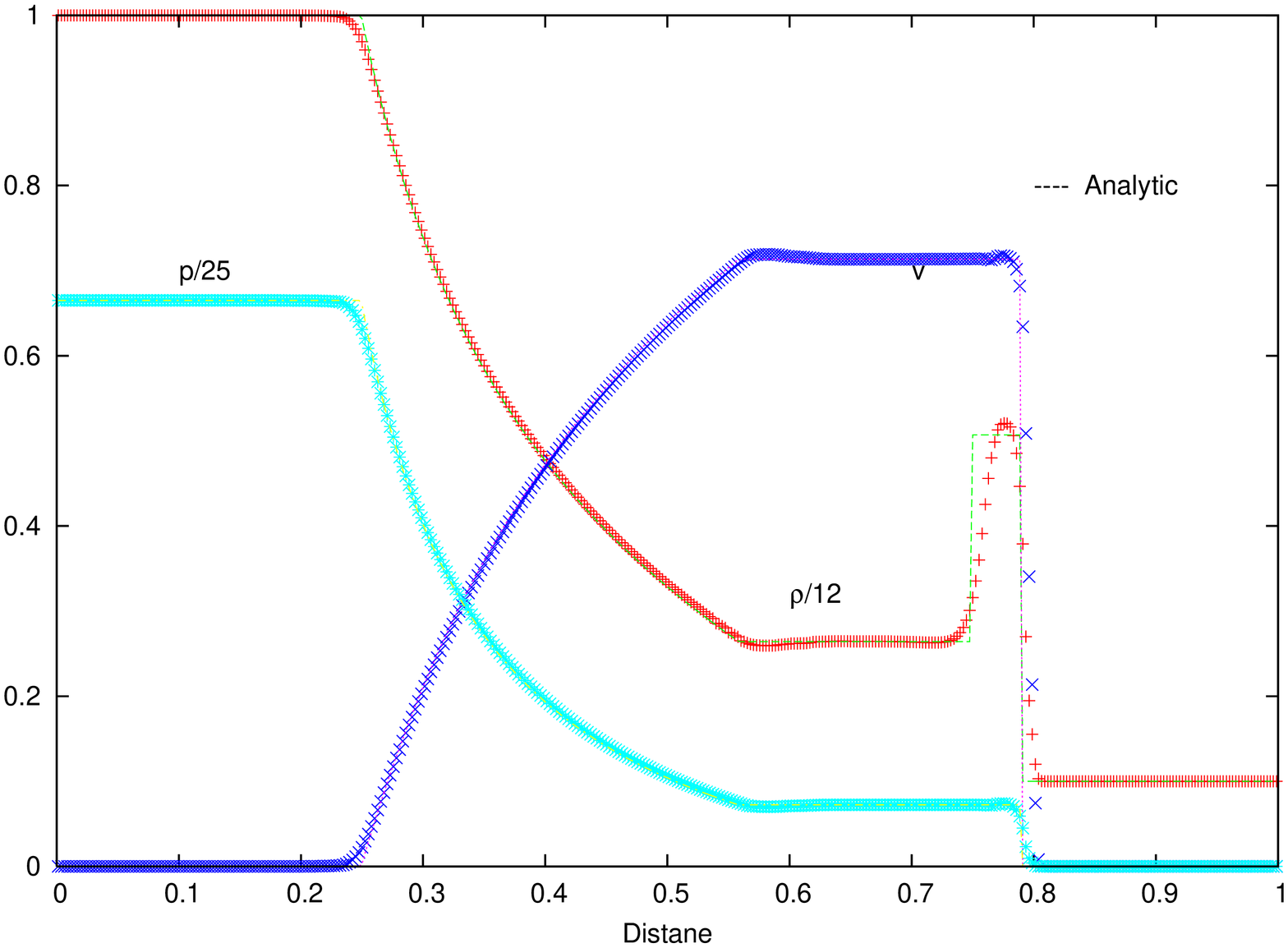}\\
  \includegraphics[angle=-90,width=0.8\textwidth]{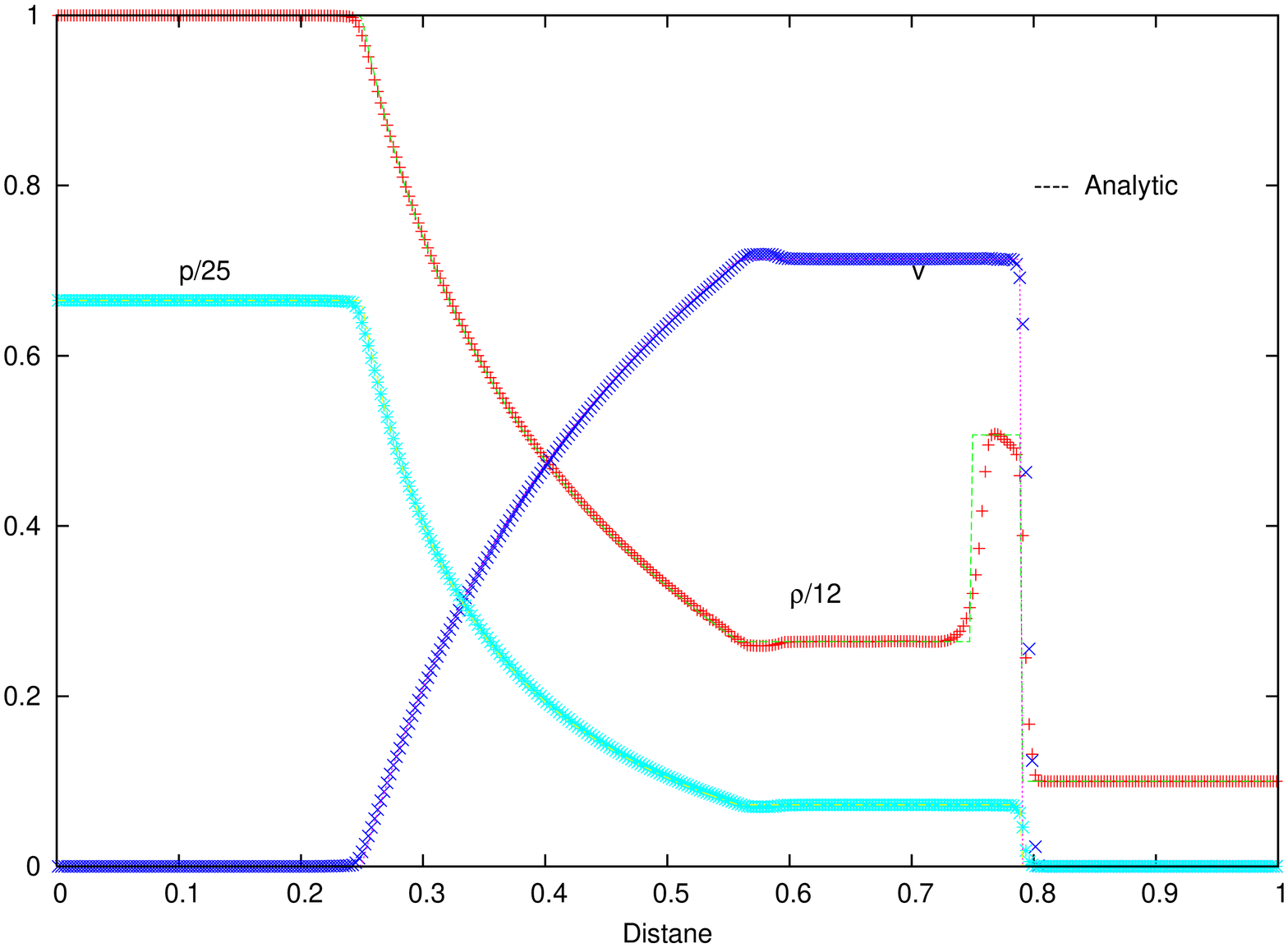}
  \caption{Density, Pressure and velocity profiles for 
test 3 (Sec. 5.1) at t=0.35 using
3$^{rd}$ (top) and 4$^{th}$ (bottom) order WENO reconstructions and n=400.}
  \label{figs5_6}
  \end{center}
\end{figure}

\begin{figure}[!h]
  \begin{center}
  \includegraphics[angle=-90,width=0.8\textwidth]{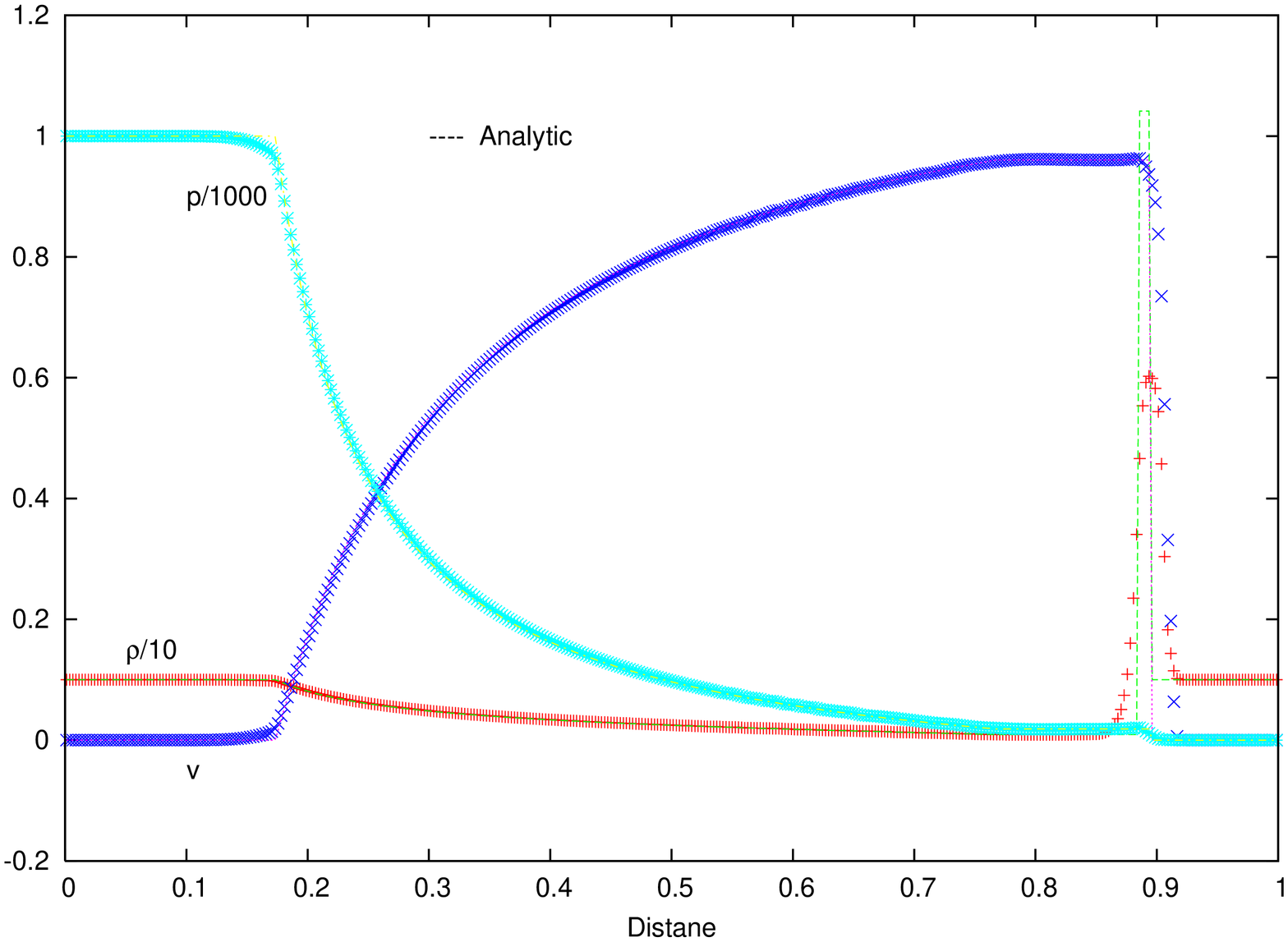}\\
  \includegraphics[angle=-90,width=0.8\textwidth]{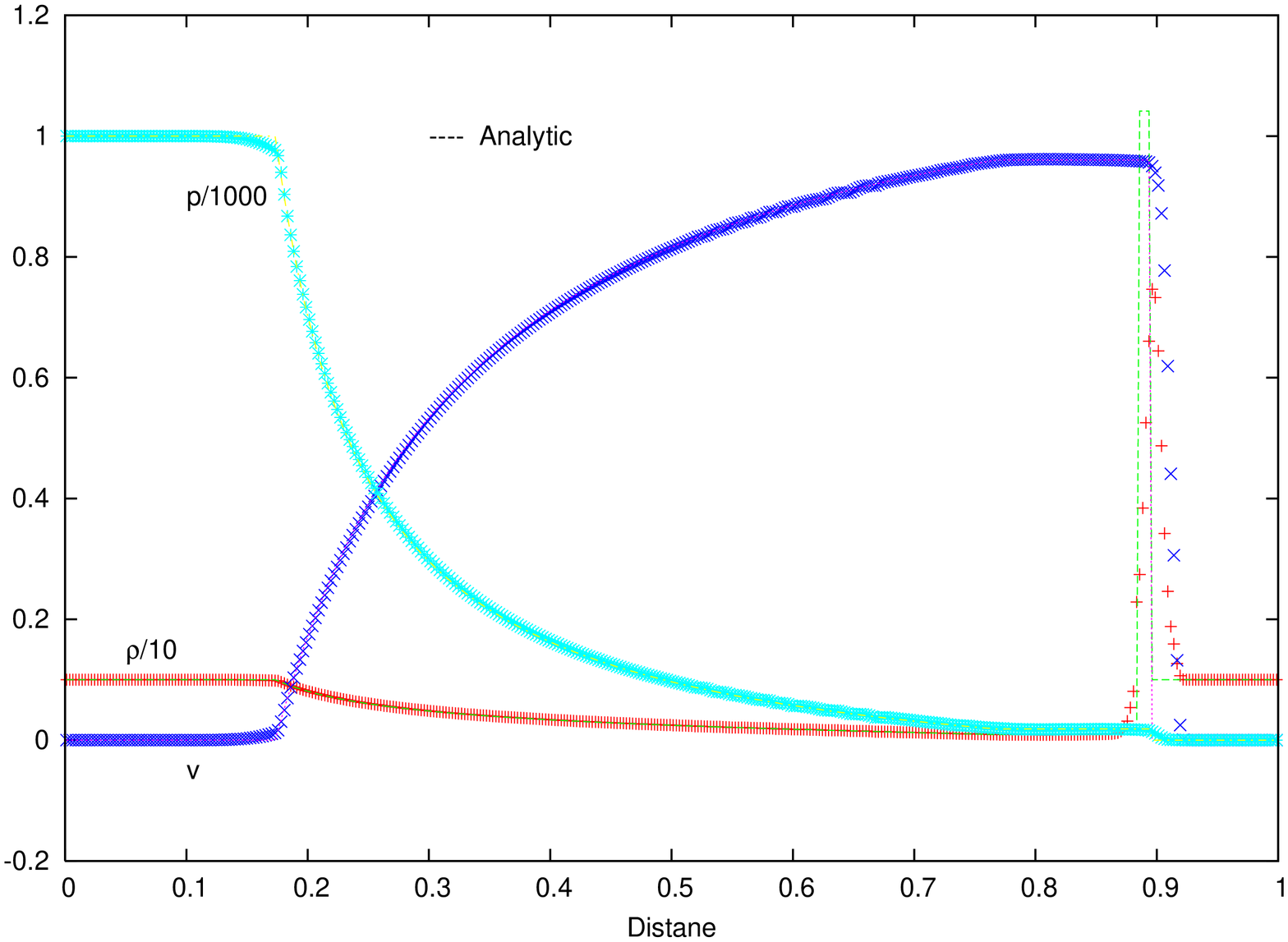}
  \caption{Density, Pressure and velocity profiles for 
test 4 (Sec. 5.1) at t=0.4 using
3$^{rd}$ (top) and 4$^{th}$ (bottom) order WENO reconstructions and n=400.}
  \label{figs7_8}
  \end{center}
\end{figure}

\begin{figure}[!h]
\begin{center}
  \includegraphics[angle=-90,width=0.8\textwidth]{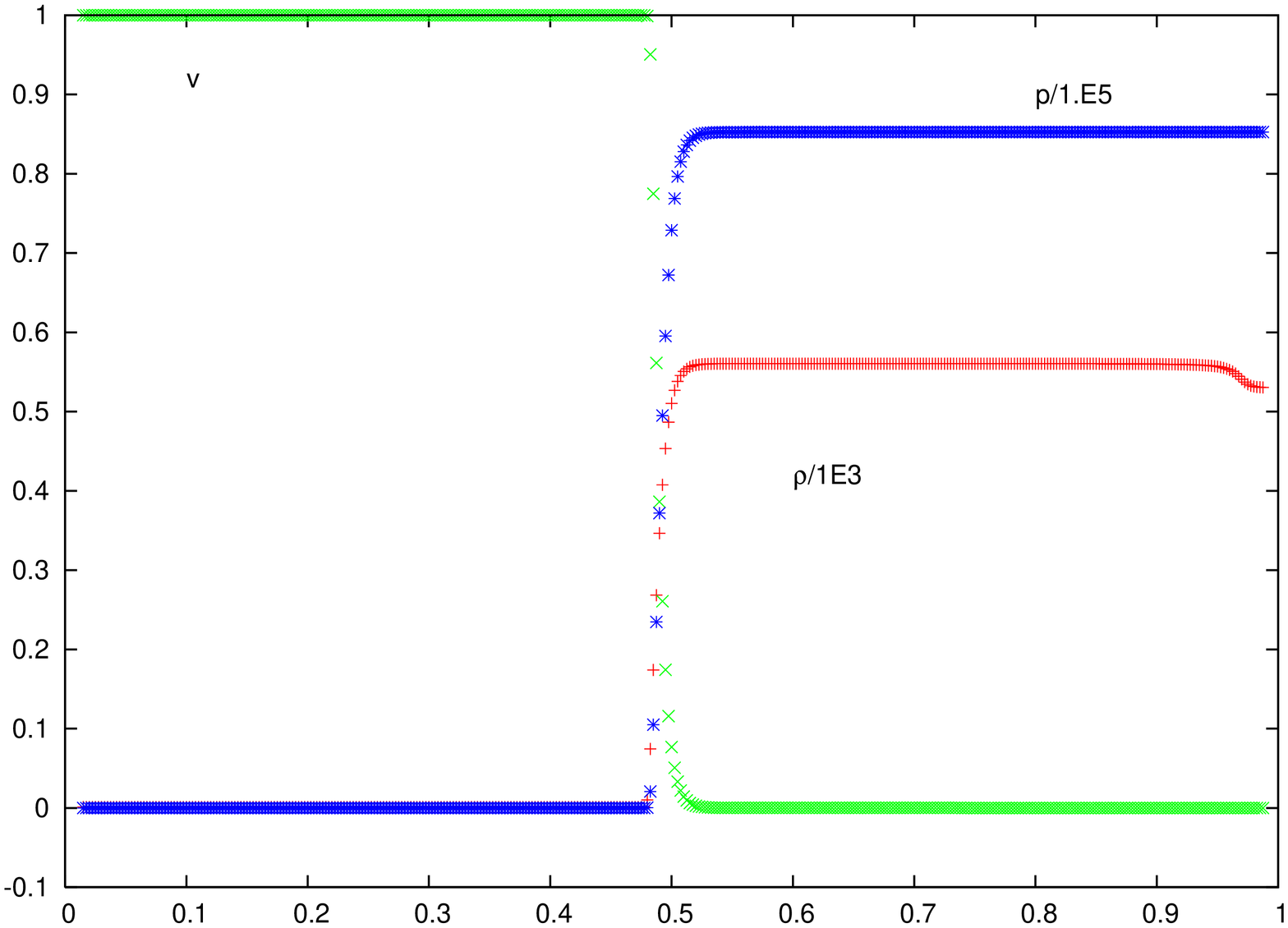}\\
  \includegraphics[angle=-90,width=0.8\textwidth]{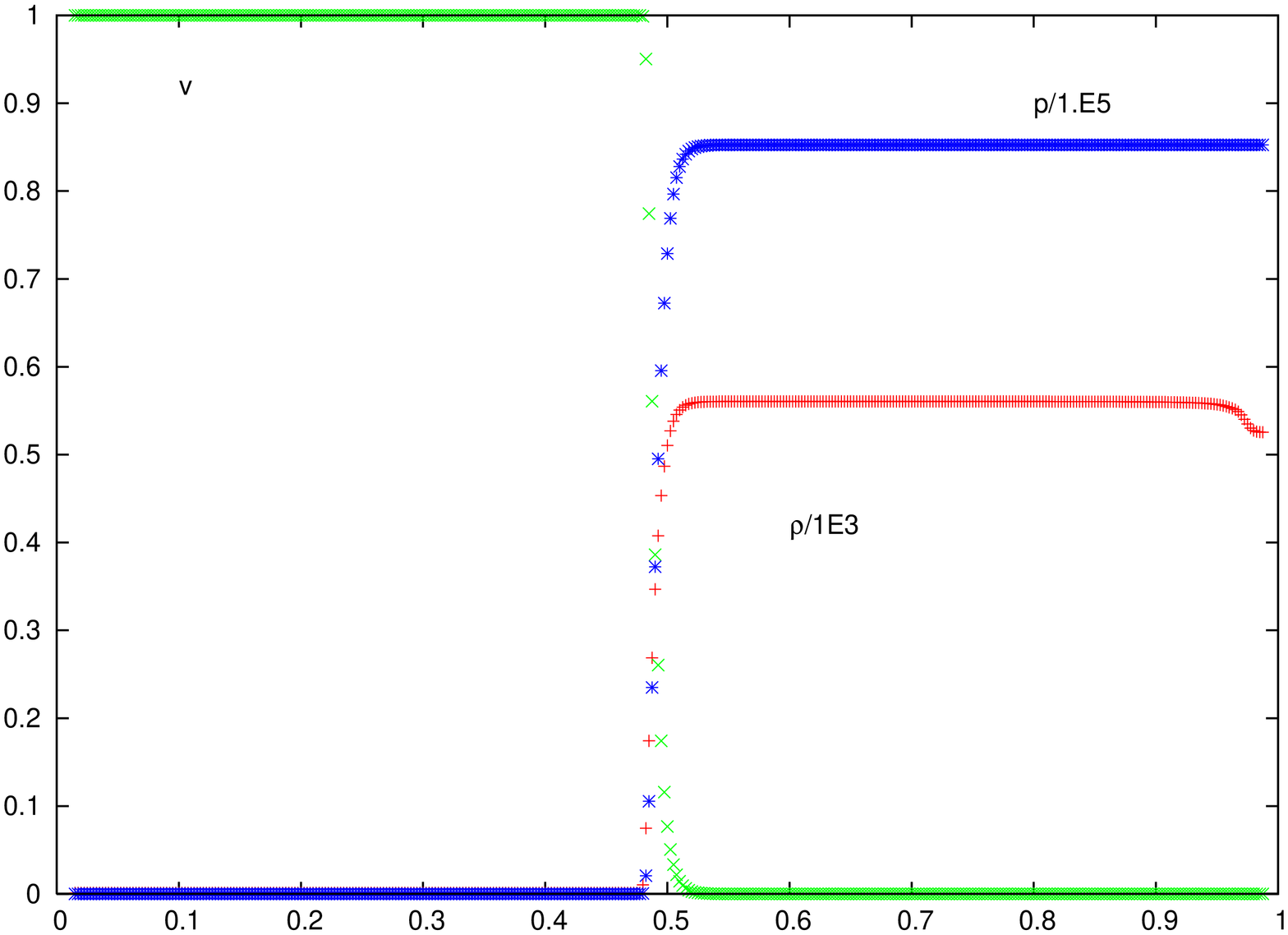}
  \caption{Density, Pressure and velocity profiles for 
test 5 (Sec. 5.1), v= .99999  using 
3$^{rd}$ (top) and 4$^{th}$ (bottom) order WENO reconstructions and n=400.}
  \label{figs9_10}
  \end{center}
\end{figure}

\begin{figure}[!h]
  \begin{center}
  \includegraphics[angle=-90,width=0.8\textwidth]
{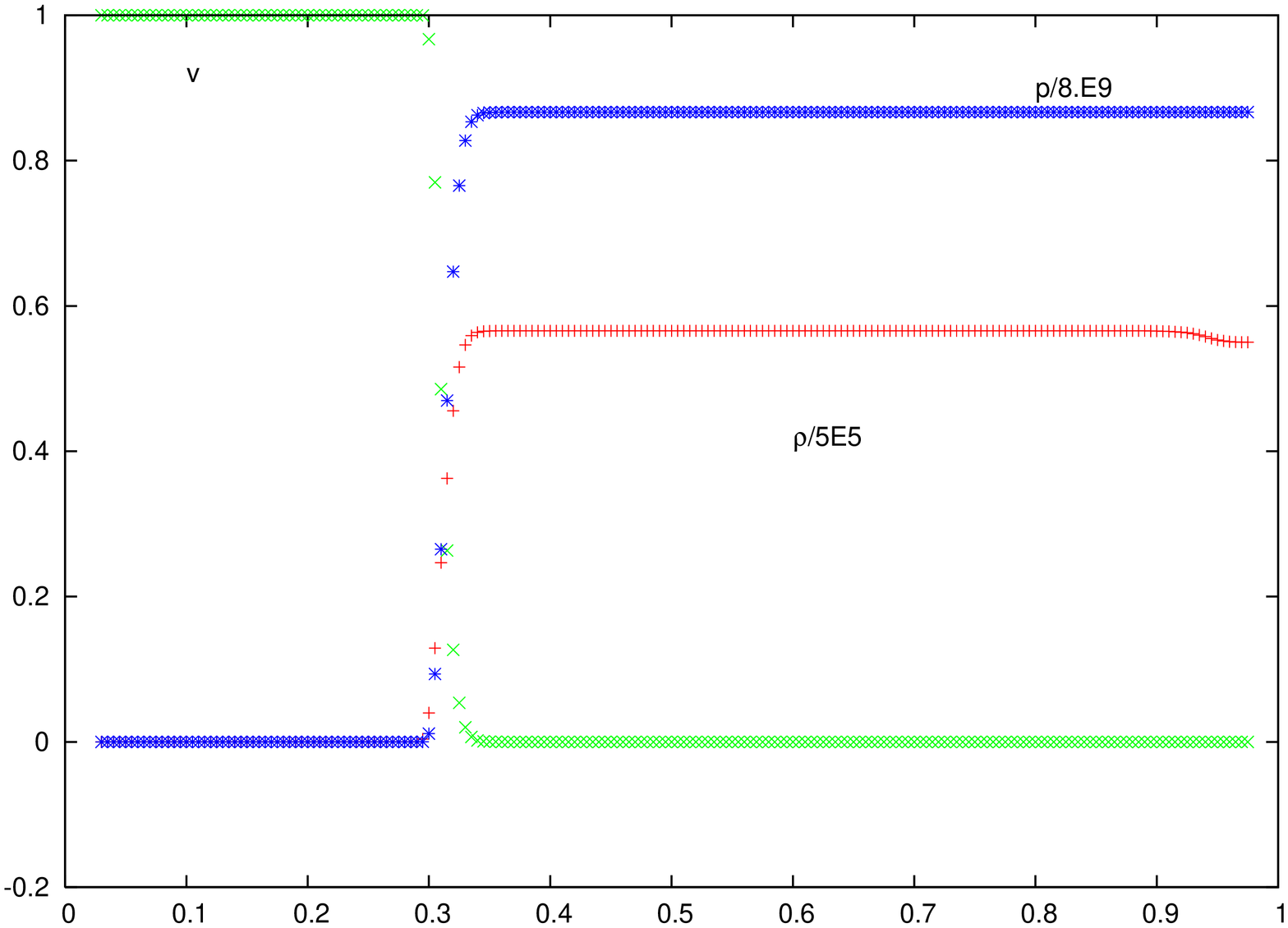}\\
  \includegraphics[angle=-90,width=0.8\textwidth]
{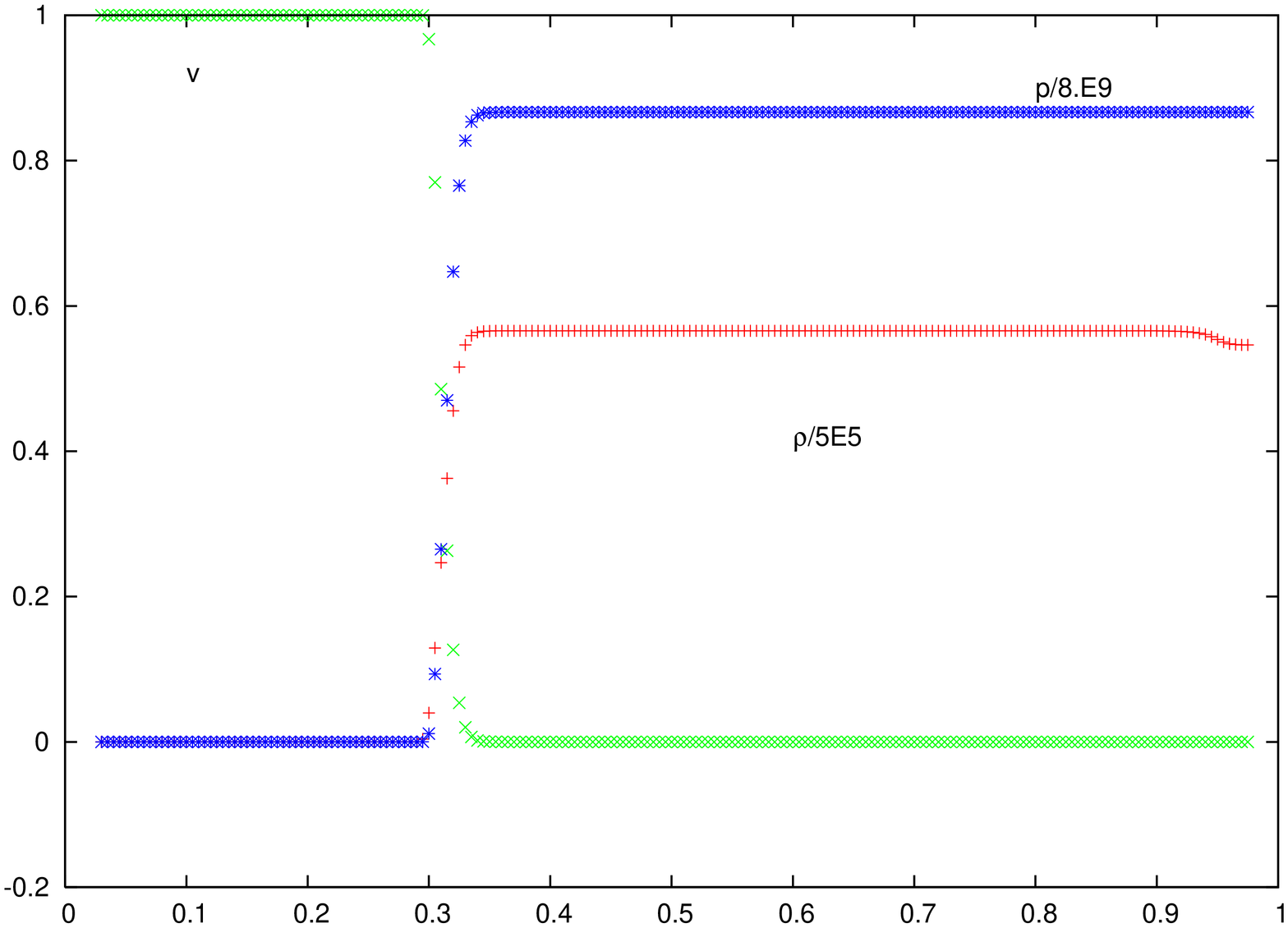}
  \caption{Density, Pressure and velocity profiles for 
test 5 (Sec. 5.1), v= .9999999999  using 
3$^{rd}$ (top) and 4$^{th}$ (bottom) order WENO reconstructions and n=400.}
  \label{figs11_12}
  \end{center}
\end{figure}
    
\begin{figure}[!h]
  \begin{center}
  \includegraphics[angle=-90,width=0.8\textwidth]{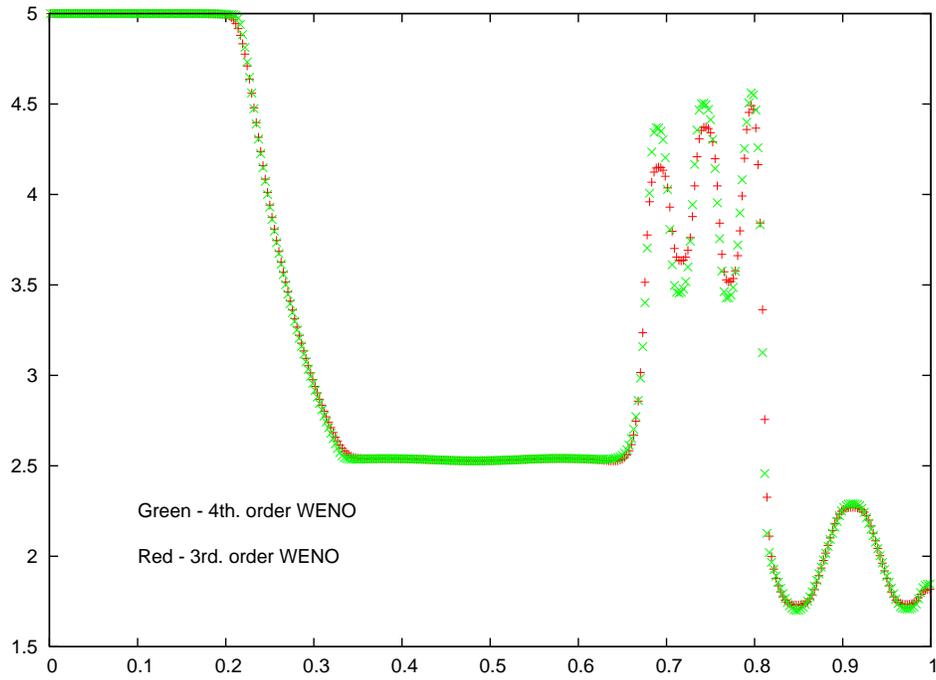}
  \caption{Mixed soulution test (test 5, Sec. 5.1) for 3$^{rd}$ and 4$^{th}$ 
order WENO reconstruction at t= .36. Shown ar the density profiles.}
  \label{figs13}
  \end{center}
\end{figure}

\begin{figure}[!h]
  \begin{center}
  \includegraphics[angle=-90,width=0.8\textwidth]{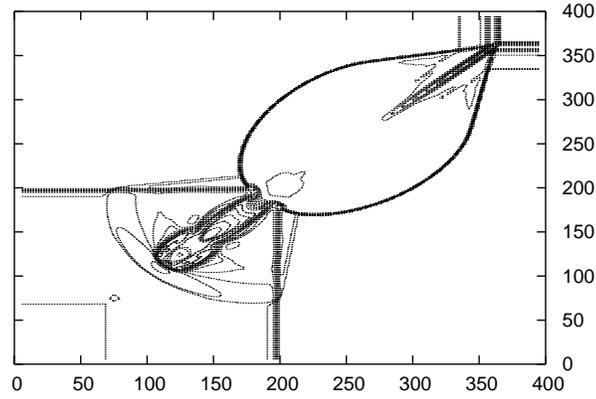} \\
  \includegraphics[angle=-90,width=0.8\textwidth]{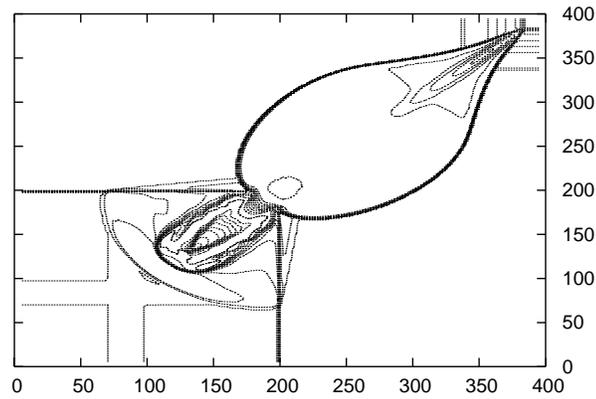} 
  \caption{2D Riemann problem, (Sec. 6.1). Logarithmic density 
plots are shown using our scheme (top) using the 4$^{th}$ order WENO schme 
and that using the PPM reconstruction method (bottom) of 
Lucas-Serano et al. (2004). 
Solution showm at t=0.4 and n=400}
  \label{}
  \end{center}
\end{figure}

\begin{figure}[!h]
  \begin{center}
  \includegraphics[angle=-90,width=1.0\textwidth]{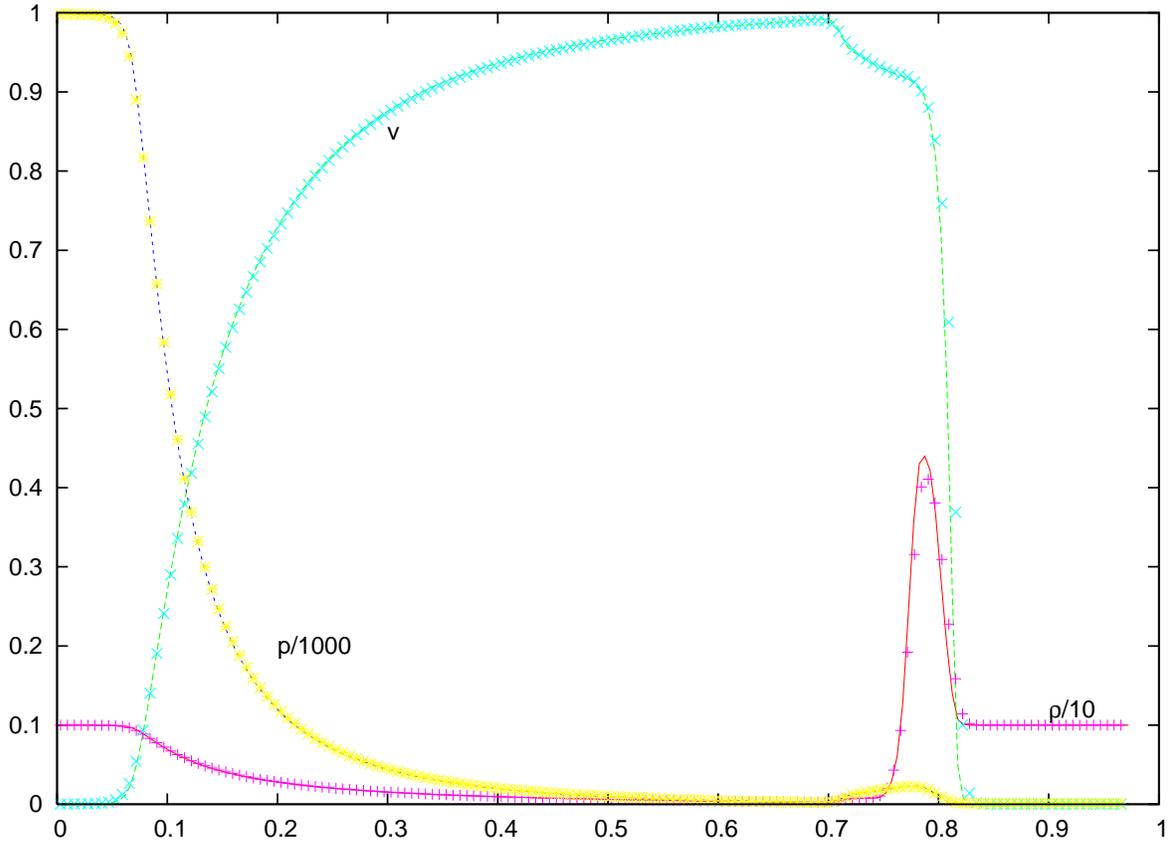}
  \caption{Cylindrically symmetric blast wave 
(Sec. 6.2) using 4$^{th}$ order WENO. Solution 
at t=0.4 and n=250.  Solid lines 
represent one dimensional calculations.}
  \label{}
  \end{center}
\end{figure}

\begin{figure}[!h]
  \begin{center}
  \includegraphics[angle=-90,width=1.0\textwidth]
{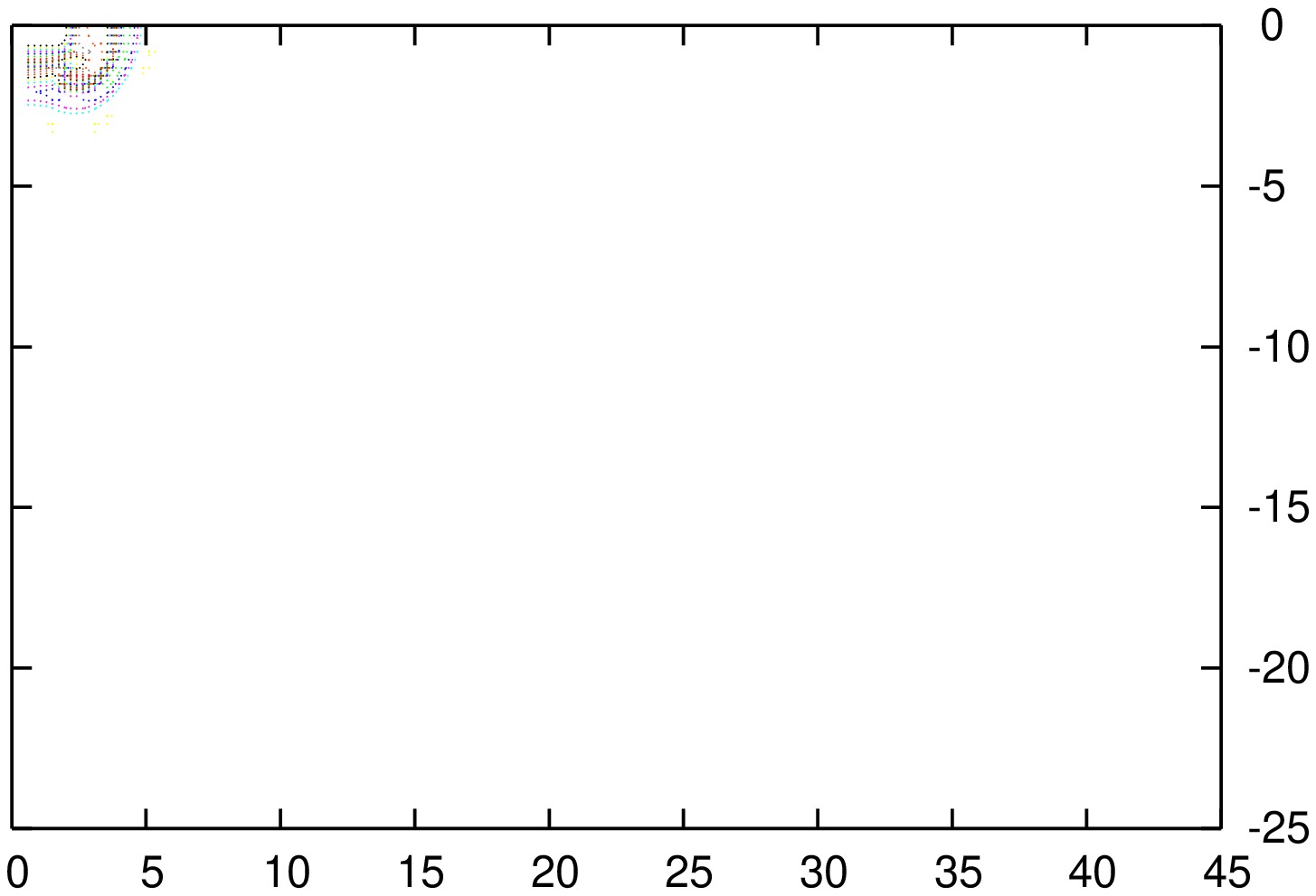}\\
  \includegraphics[angle=-90,width=1.0\textwidth]
{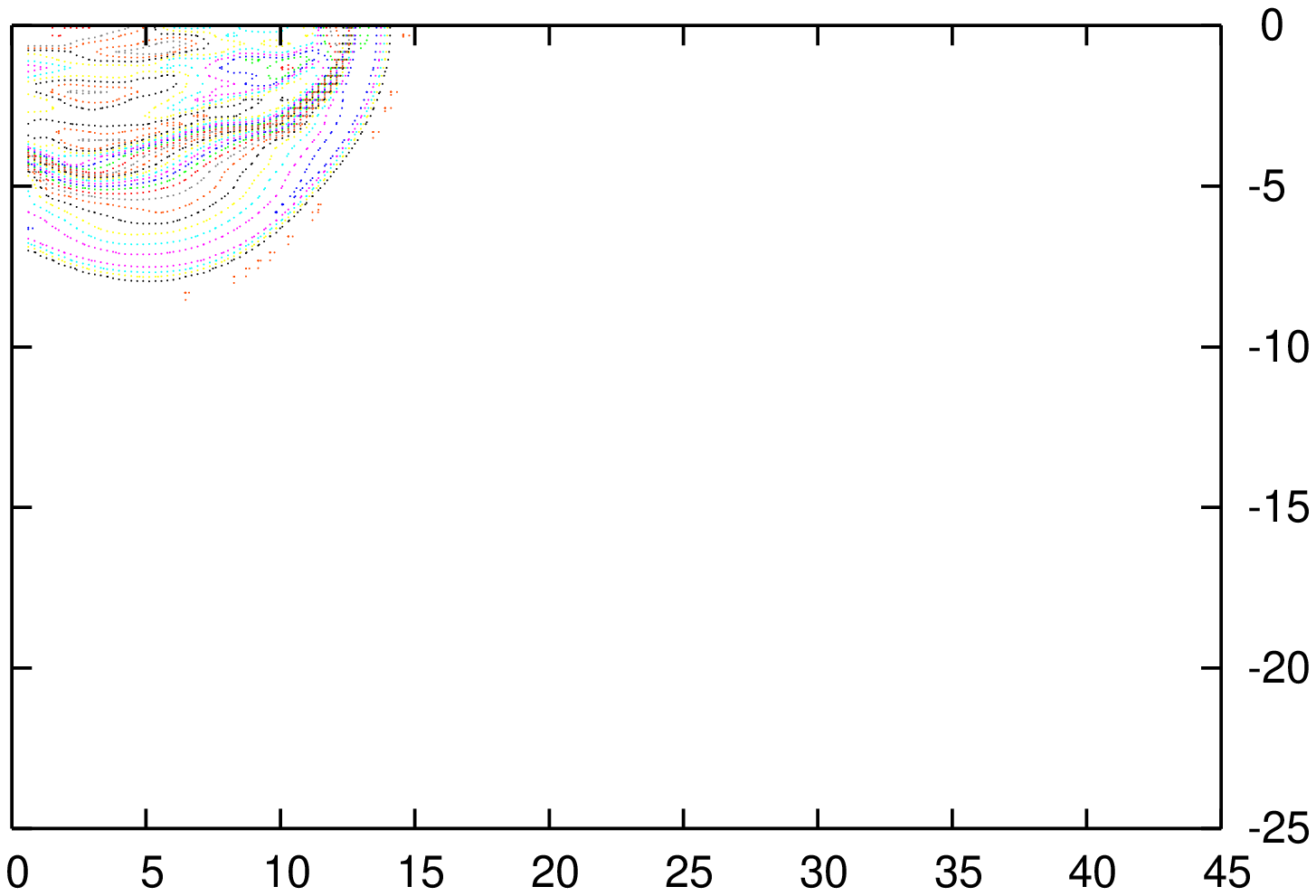}
  \caption{Two dimensional jet problem (Sec. 6.3). 
Density profile at t=5 (top), t=25 (bottom)}
  \label{}
  \end{center}
\end{figure}

\begin{figure}[!h]
  \begin{center}
  \includegraphics[angle=-90,width=1.0\textwidth]
{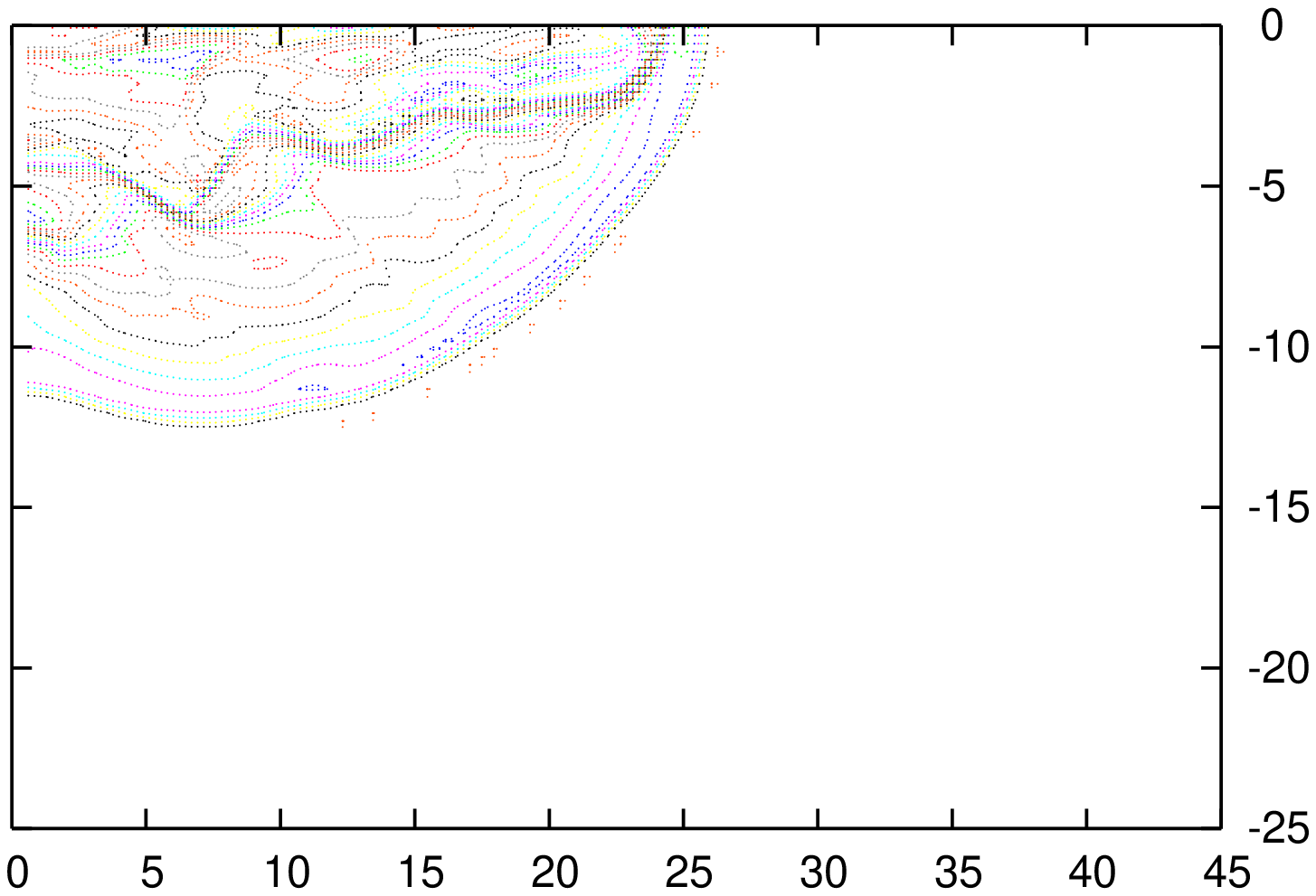}\\
  \includegraphics[angle=-90,width=1.0\textwidth]
{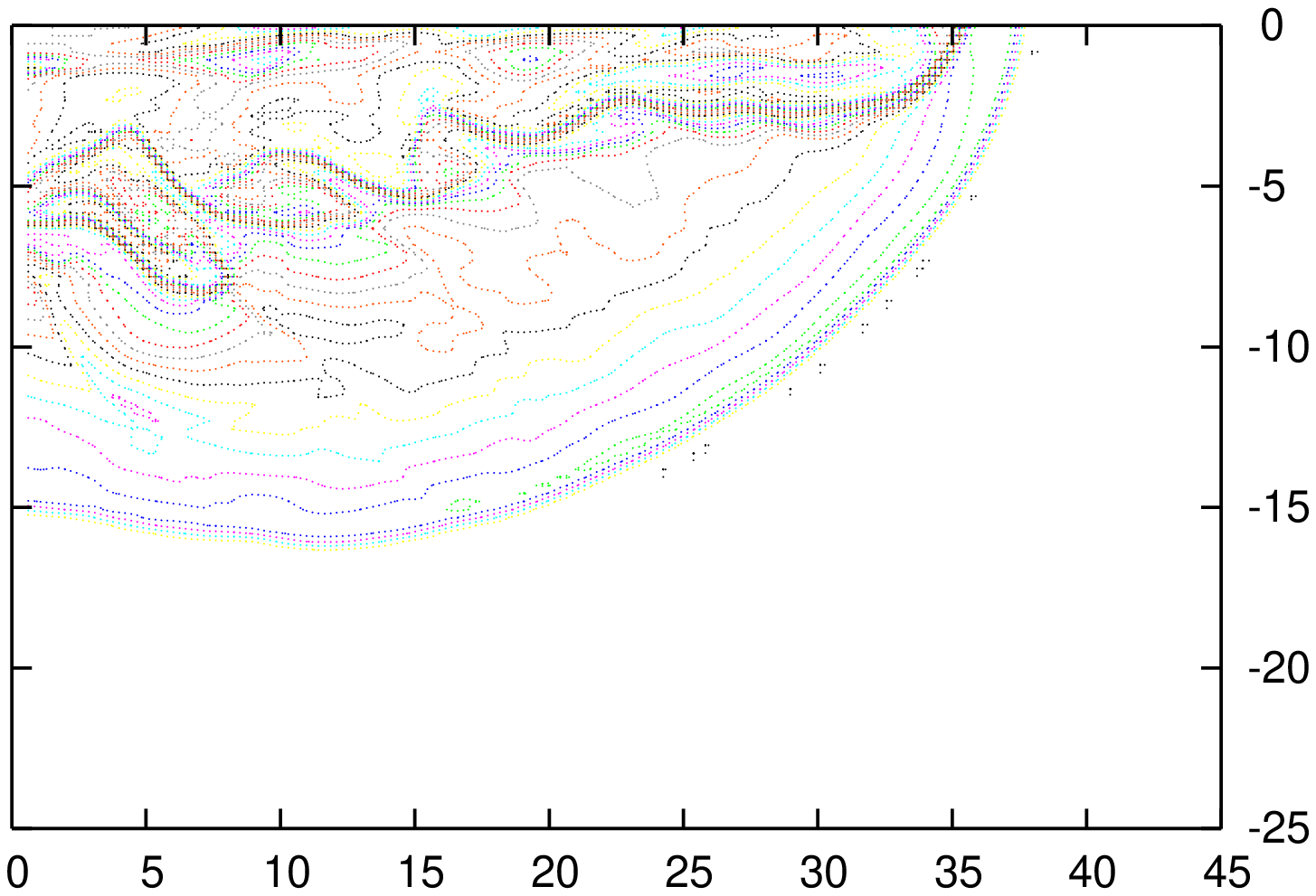} 
  \caption{Two dimensional jet problem (Sec. 6.3). 
Density profile at t=50 (top), t=70 (bottom)}
  \label{}
  \end{center}
\end{figure}

\begin{figure}[!h]
  \begin{center}
  \includegraphics[angle=-90,width=1.0\textwidth]
{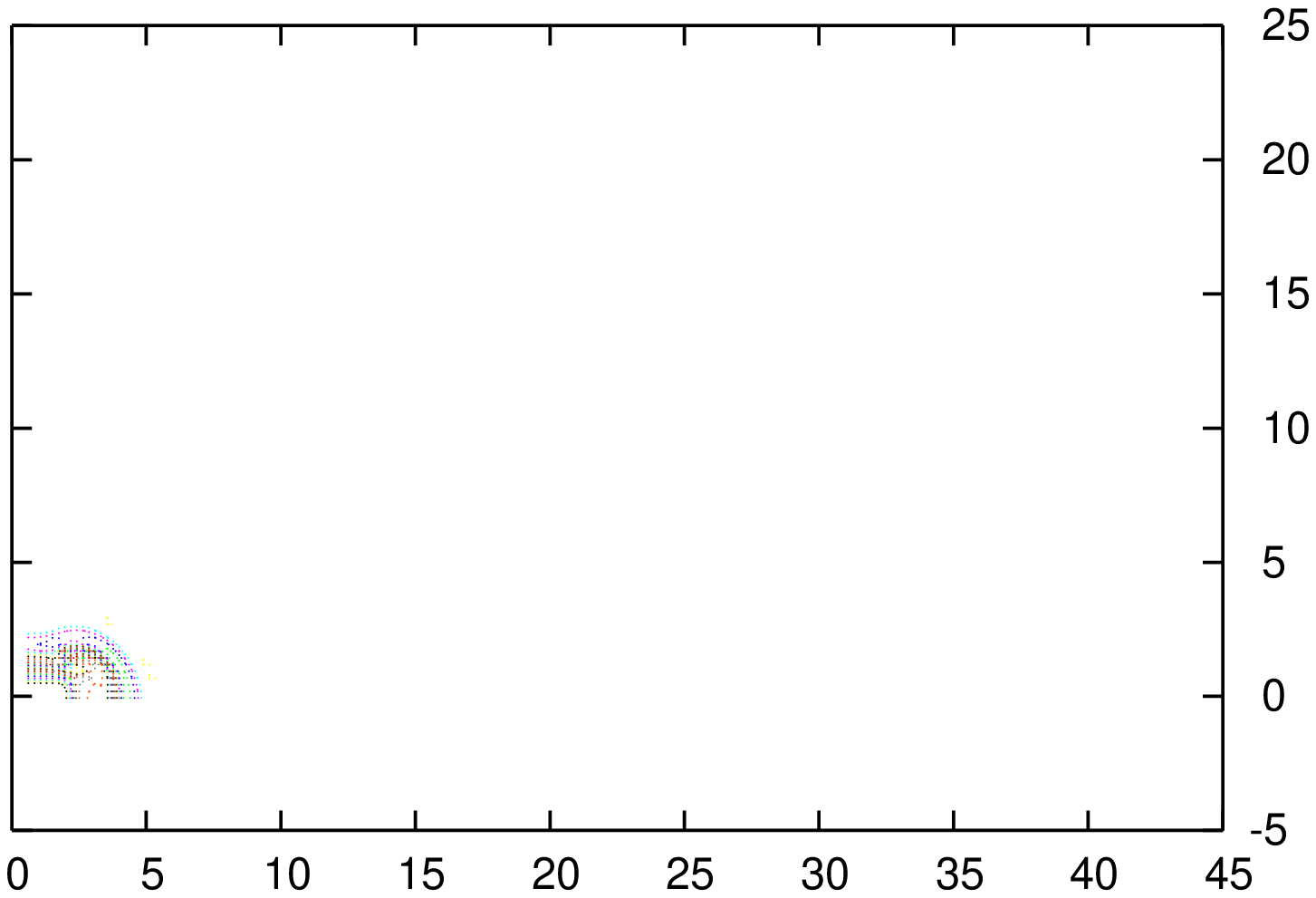}\\
  \includegraphics[angle=-90,width=1.0\textwidth]
{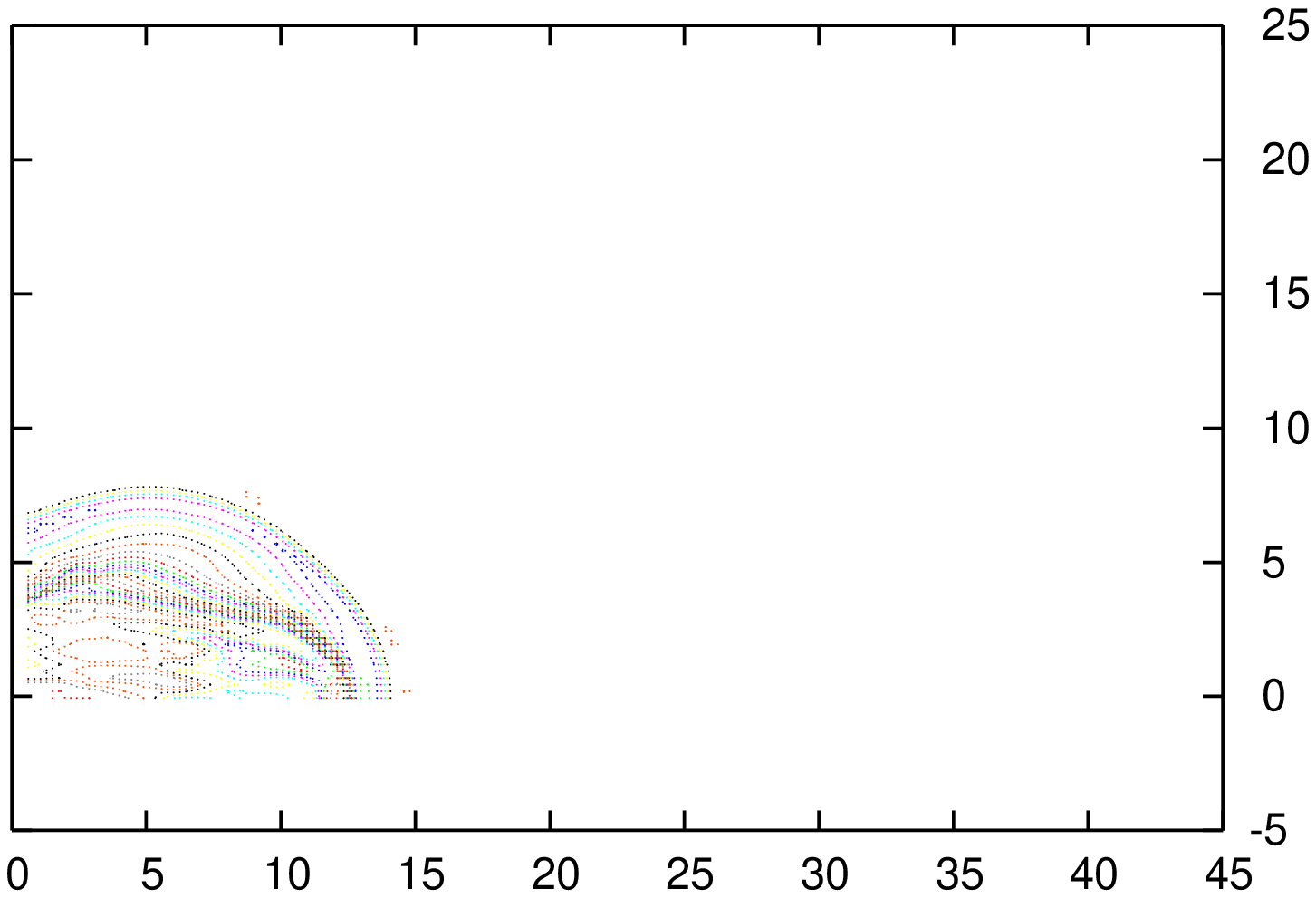}
  \caption{Two dimensional jet problem (Sec. 6.6.3). 
Density profile at t=5 (top), t=25 (bottom)}
  \label{}
  \end{center}
\end{figure}

\begin{figure}[!h]
  \begin{center}
  \includegraphics[angle=-90,width=1.0\textwidth]
{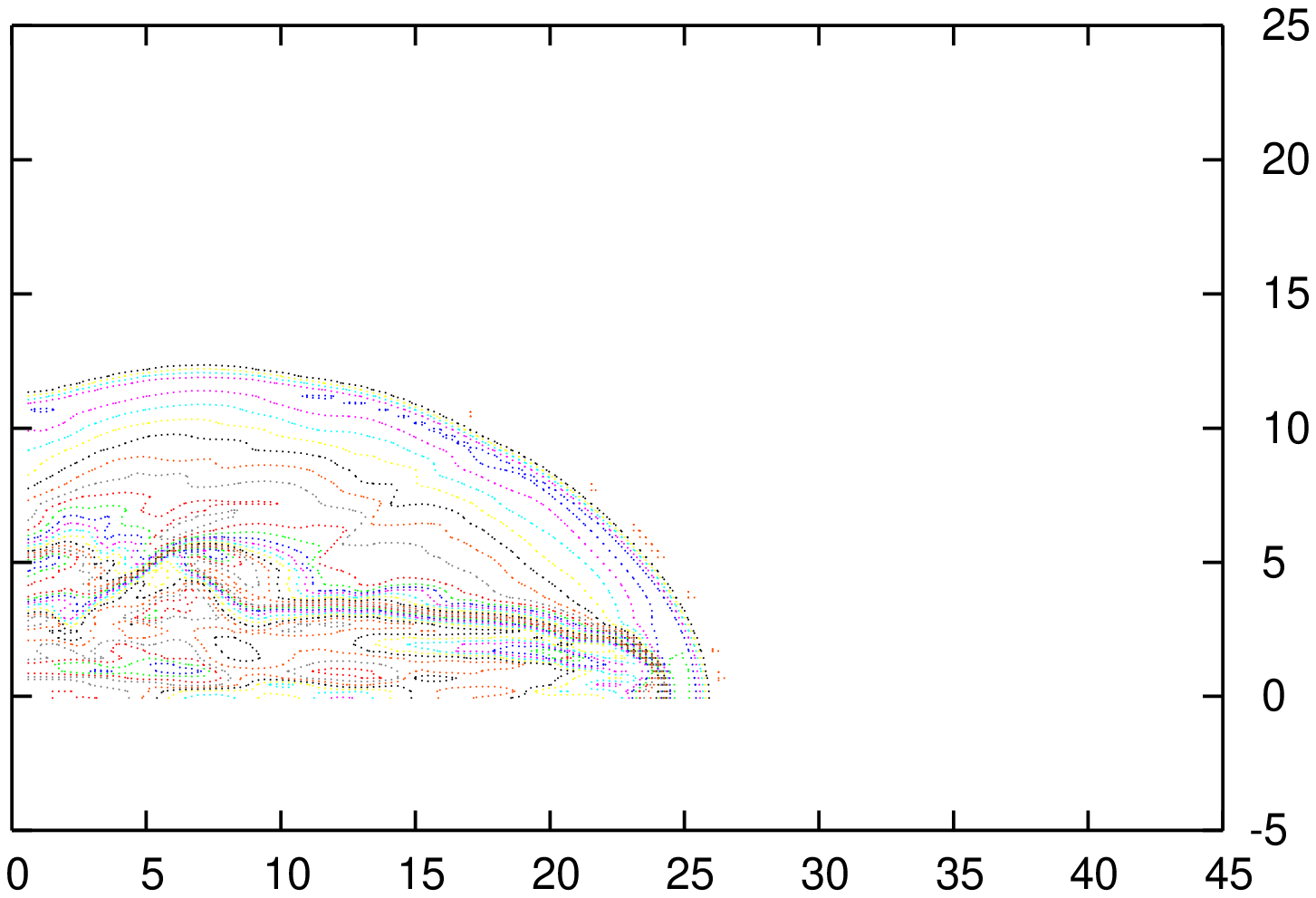}\\
  \includegraphics[angle=-90,width=1.0\textwidth]
{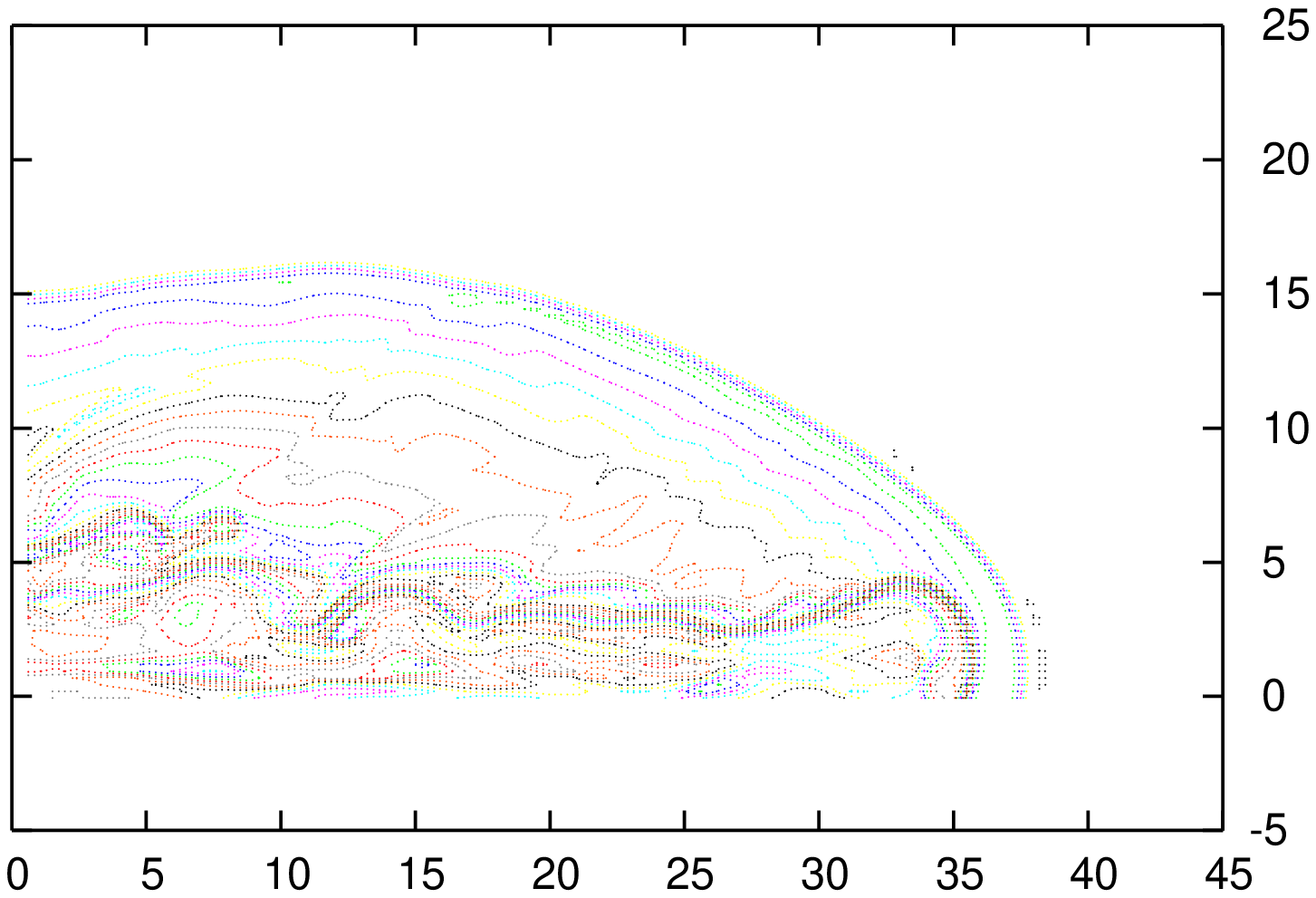} 
  \caption{Two dimensional jet problem (Sec. 6.3). 
Density profile at t=50 (top), t=70 (bottom)}
  \label{}
  \end{center}
\end{figure}

\end{document}